\documentclass{aa}
\usepackage{longtable}
\usepackage{supertabular}
\usepackage{graphics}
\usepackage{rotating}
\usepackage{natbib}
\begin{document}
\thesaurus{0.8(0.9.10.1; 0.9.13.2; 0.8.16.5)}
\title{RASS-SDSS Galaxy Clusters Survey.} 
\subtitle{III. Scaling relations of galaxy clusters}
\author{P. Popesso\inst{1}, A. Biviano\inst{2}, H. B\"ohringer\inst{1}, M. Romaniello\inst{3}, W. Voges\inst{1}}
\institute{ Max-Planck-Institut fur extraterrestrische Physik, 85748 Garching, Germany
\and INAF - Osservatorio Astronomico di Trieste, via G. B. Tiepolo 11, I-34131, Trieste, Italy
\and European Southern Observatory, Karl Scharzschildstr. 2, Garching b. M\"unchen, Germany}
\authorrunning{P. Popesso et al.}
\maketitle

\begin{abstract}
We use the RASS-SDSS galaxy cluster sample to compare the quality of
optical and X-ray luminosities as predictors of other cluster
properties such as their masses, temperatures, and velocity
dispersions. We use the SDSS spectroscopic data to estimate the
velocity dispersions and the virial masses of a subsample of 69
clusters within $r_{500}$ and $r_{200}$.  The ASCA temperature of the
intra-cluster medium, $T_X$, is retrieved from the literature for a
subsample of 49 clusters. For this subsample we estimate the cluster
masses also by using the mass-temperature relation. We show that the
optical luminosity, $L_{op}$, correlates with the cluster mass much
better than the X-ray luminosity, $L_X$. $L_{op}$ can be used to
estimate the cluster mass with an accuracy of 40\% while $L_X$ can
predict the mass only with a 55\% accuracy. We show that correcting
$L_X$ for the effect of a cool core at the center of a cluster, lowers
the scatter of the $L_X-M$ relation only by 3\%. We find that the
scatter observed in the $L_{op}-L_X$ relation is determined by the
scatter of the $L_X-M$ relation. The mass-to-light ratio in the SDSS
$i$ band clearly increases with the cluster mass with a slope
$0.2\pm0.08$. The optical and X-ray luminosities correlate in excellent
way with both $T_X$ and $\sigma_V$ with an orthogonal scatter of 20\%
in both relations. Moreover, $L_{op}$ and $L_X$ can predict with the
same accuracy both variables.  We conclude that the cluster optical
luminosity is a key cluster parameter since it can give important
information about fundamental cluster properties such as the mass, the
velocity dispersion, and the temperature of the intra-cluster medium.
\end{abstract}

\section{Introduction}
Clusters of galaxies are the most massive gravitationally bound systems
in the universe.  The mass is the most important property of these
systems. The cluster mass function and its evolution provide
constraints on the evolution of large-scale structure and important
cosmological parameters such as $\Omega_m$ and $\sigma _8$. The
mass-to-light ratio of clusters provides one of the most robust
determination of $\Omega _m$ in connection with the observed light
density in the Universe via the Oort (1958) method. For these reasons,
over the last 70 years (starting with Zwicky 1933, 1937, and Smith
1936), much effort has been spent measuring the mass of clusters using
a number of techniques. These include: (i) dynamical methods applied
on the galaxy distributions derived from redshift surveys, or (ii)
based on the distribution and temperature of the diffuse hot gas in
the intracluster medium (ICM), observed at X-ray wavelength, (iii)
gravitational lensing, and (iv) observations of the Sunyaev-Zeldovich
effect. All these methods are in general quite expensive in terms of
the observational resources required, especially for high redshift
clusters.
 
In the literature several comparisons have been made between the
different methods for determining the mass of the clusters. These
include comparisons between the X-ray and strong-lensing methods (see,
e.g., Wu 2000) in cores of clusters, between X-ray and weak lensing
methods (see, e.g., Smail et al. 1997), and between the dynamical and
the X-ray methods (see, e.g., Wu 2000). In particular, Girardi et
al. (1998) and Rines et al. (2003) have shown that consistent results
can be obtained between these last two methods.
 
The existence of a fundamental plane for the global properties of
galaxy clusters (Schaeffer et al. 1993; Adami et al.  1998; Fujita \&
Takahara 1999) naturally implies that other properties, such as
cluster luminosities, velocity dispersions, X-ray temperatures, can be
used to infer the cluster masses. It is not known whether it is the
X-ray or the optical luminosity that correlates better with the
cluster mass. Reiprich \& B\"ohringer (2002, R02 hereafter) showed
that a tight correlation exists between the X-ray total luminosity of
the clusters and the mass with a scatter of 60\%. Girardi et
al. (2000, 2002) analysed the relation between mass and optical
luminosity in the blue band, and detetrmined the mass-to-light ratio,
for a sample of 162 clusters using inhomogeneous photometric data. Yee
\& Ellingson (2003) analysed a sample of 16 X-ray luminous clusters
from the {\em Canadian Network for Observational Cosmology} (CNOC)
survey. They used the cluster optical richness, rather than the
optical luminosity, and showed that it is well correlated with other
global properties of galaxy clusters as their velocity dispersion, their
intracluster gas temperature, and their total mass. Lin et
al. (2003) analysed a sample of 27 clusters with near-infrared data
from the Two Micron All Sky Survey (2MASS) and available X-ray
temperature. They showed that the $K$-band luminosity of a cluster can
be used to estimate its mass with 45\% accuracy.
 
Analysing the relation between optical luminosity and cluster mass is
not an easy task. The lack of optical wide field surveys in the past
did not allow to measure in the proper way the optical luminosity in
galaxy systems. Until now the uncertainties in luminosity
determination came from the corrections for the calibration of
inhomogenous photometric data, background galaxy contamination and the
need to extrapolate the sum of measured luminosities of galaxy members
to include faint galaxies and the outer parts of the systems, beyond
the region studied. The use of the {\em Sloan Digital Sky Survey} (SDSS) for
the optical data allows us to overcome all the problems related to the
optical luminosity estimation.

In this paper we use the RASS-SDSS galaxy cluster sample (Popesso et
al. 2004) to study the correlations between the optical luminosity and
other important properties of galaxy clusters such as their mass,
line-of-sight velocity dispersion, temperature and X-ray
luminosity. Moreover, the RASS-SDSS galaxy cluster sample allows us to
compare the correlations of the optical and, respectively, the X-ray
luminosity with other global properties of the systems. The excellence
of the second release of the SDSS (SDSS-DR2, Abazajian et al.  2004)
in terms of its size, depth and sky coverage, the accurate photometry
in 5 different optical wavebands, and the detailed spectroscopy for
more the 260,000 galaxies, give us unprecedented advantages in
comparison to the previous studies.  Firstly, the sky coverage (3324
$\rm{deg}^2$) gives us the possibility of studying a large sample of
clusters with completely homogeneous photometric data. Secondly, the
accurate estimation of the spectroscopic redshift for a large
subsample of galaxies allows us to define an accurate membership for
any cluster and, thus, perform a detailed dynamical analysis of the
system within the same survey. Thirdly, the sky coverage of the survey
allows also to overcome the well known problem of the statistical
subtraction of the galaxy background. We use large areas of the survey
to define a mean global galaxy background and a region close to the
clusters to determine the local galaxy background in order to check
for systematics in the field subtraction. Finally, the apparent
magnitude of the SDSS DR2 in all the five bands is sufficiently deep
(e.g.  $r_{lim}=22.2$, 95\% completeness) that, at the mean redshift
of our cluster sample ($z\sim 0.10$), the cluster luminosity function
(LF) is sampled down to a significant part of the faint end. Moreover,
the use of the {\em ROSAT All Sky Survey} (RASS) allows us to define also
the X-ray properties in an homogeneous way for all the systems and
perform a detailed comparison with the optical luminosity.
 
In this paper we show that the optical luminosity, $L_{op}$, is
strongly correlated with the other global properties of a galaxy
cluster, allowing its use as an estimator for quantities such as the
velocity dispersion, the mass, and the intra-cluster gas temperature,
$T_X$. We demonstrate that $L_{op}$ is a better predictor of the
virial mass than the X-ray luminosity, which makes $L_{op}$ an
important defining parameter of galaxy clusters, and an extremely
useful cosmological tool.  Throughout this paper, we use $H_0=70$ km
s$^{-1}$ Mpc$^{-1}$ in a flat cosmology with $\Omega_0=0.3$ and
$\Omega_{\Lambda}=0.7$ (e.g. Tegmark et al. 2004).

\section{The cluster sample and the data}
The ROSAT-SDSS galaxy cluster catalog comprises 130 systems detected
in the ROSAT All Sky Survey (RASS).  The X-ray cluster properties and
the redshifts have been taken from different catalogs of X-ray
selected clusters: the ROSAT-ESO flux limited X-ray cluster sample
(REFLEX, B\"ohringer et al.  2001, 2002), the Northern ROSAT All-sky cluster
sample (NORAS, B\"ohringer et al.  2000), the NORAS 2
cluster sample (Retzlaff 2001), the ASCA Cluster Catalog (ACC) from
Horner et al. (2001) and the Group Sample (GS) of Mulchaey et
al. (2003).

The optical photometric data are taken from the SDSS DR2 ( Fukugita
1996, Gunn et al. 1998, Lupton et al. 1999, York et al. 2000, Hogg et
al. 2001, Eisenstein et al. 2001, Smith et al. 2002, Strauss et
al. 2002, Stoughton et al.  2002, Blanton et al. 2003 and Abazajian et
al.  2003).  The SDSS consists of an imaging survey of $\pi$
steradians of the northern sky in the five passbands $u, g, r ,i, z,$
in the entire optical range from the atmospheric ultraviolet cutoff in
the blue to the sensitivity limit of silicon in the red.  The survey
is carried out using a 2.5 m telescope, an imaging mosaic camera with
30 CCDs, two fiber-fed spectrographs and a 0.5 m telescope for the
photometric calibration.  The imaging survey is taken in drift-scan
mode.  The imaging data are processed with a photometric pipeline
(PHOTO, Lupton et al. 2001) specially written for the SDSS data.  For
each cluster we defined a photometric galaxy catalog as described in
Section 3 of Popesso et al. (2004, see also Yasuda et al. 2001).

For the analysis in this paper we use only SDSS Model magnitudes. Due
to a bug of PHOTO, found during the completion of DR1, the model
magnitudes were systematically under-estimated by about 0.2-0.3
magnitudes for galaxies brighter than 20th magnitude, and accordingly
the measured radii were systematically too large. This problem has
been fixed in the SDSS DR2, therefore the model magnitude can be
considered a good estimate of the galaxy total luminosity at any
magnitude and is not dependent on the seeing as the Petrosian
magnitudes.

The spectroscopic component of the survey is carried out using two
fiber-fed double spectrographs, covering the wavelength range
3800--9200 \AA, over 4098 pixels. They have a resolution
$\Delta\lambda/\lambda$ varying between 1850 and 2200, and together
they are fed by 640 fibers, each with an entrance diameter of 3
arcsec. The fibers are manually plugged into plates inserted into the
focal plane; the mapping of fibers to plates is carried out by a
tiling algorithm (Blanton et al. 2003) that optimizes observing
efficiency in the presence of large-scale structure. The finite
diameter of the fiber cladding prevents fibers on any given plate from
being placed closer than 55 arcsec apart.  For any given plate, a
series of fifteen-minute exposures is carried out until the mean
signal to noise ratio (S/N) per resolution element exceeds 4 for
objects with fiber magnitudes (i.e., as measured through the 3
aperture of the fiber) brighter than $g = 20.2$ and $i = 19.9$, as
determined by preliminary reductions done at the observing site. Under
good conditions (dark, clear skies and good seeing), this typically
requires a total of 45 minutes of exposure.

To create a homogeneous catalog of X-ray cluster properties, we have
computed the X-ray luminosity using only RASS data for all clusters in
the sample. The X-ray luminosity has been calculated with the
growth curve analysis (GCA) method used for the NORAS and REFLEX
cluster surveys (B\"ohringer et al. 2000) based on the RASS3 data base
(Voges et al. 1999). The GCA method is optimized for the detection of
the extended emission of clusters by assessing the plateau of the
background subtracted cumulative count rate curve. We use as final
result the total flux inside the radius $r_{200}$ which is corrected
for the missing flux estimated via the assumption of a standard
$\beta$-model for the X-ray surface brightness (see B\"ohringer et
al. 2000 for more details). The correction is typically only $8 -
10\%$ illustrating the high effectiveness of the GCA method to sample
the flux of extended sources. For a subsample of 49 galaxy clusters
we have also compiled from the literature the ASCA temperature of the
ICM.

\section{The cluster optical luminosity}
The total optical luminosity of a cluster has to be computed after
the subtraction of the foreground and background galaxy
contamination. We consider two different approaches to the
statistical subtraction of the galaxy background. We compute the
local background number counts in an annulus around the cluster and a
global background number counts from the mean of the magnitude number
counts determined in five different SDSS sky regions, each with an
area of 30 $\rm{deg^2}$. In our analysis we show the results obtained
using the optical luminosity estimated with the second method. The
optical luminosity is then computed following the prescription of
Popesso et al. (2004). The reader is referred to that paper for a detailed
discussion about the comparison between optical luminosities
calculated with different methods.

\begin{figure}
\begin{center}
\begin{minipage}{0.5\textwidth}
\resizebox{\hsize}{!}{\includegraphics{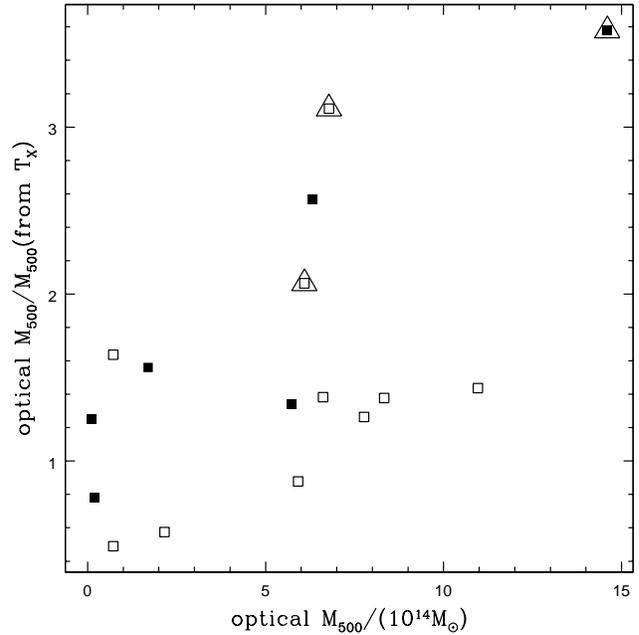}}
\end{minipage}
\end{center}
\caption{ 
Comparison of $M_{500}$ estimated from X-ray data with the mass
estimated from the dynamical analysis of the optical data. We know the
temperature for 16 clusters of the subsample with known optical
mass. For 10 of them the measure of the X-ray mass is given by R02. We
compare the the optical mass with the X-ray mass of R02, when
available, and with the mass derived from the $M-T_X$ relation
(eq. \ref{eq1}) in the other case. The empty squares are the ratio
between the optical mass and the mass given by R02. The filled
squares are the ratio between the optical mass and the mass derived
from the $M-T_X$ relation (eq. \ref{eq1}). The squares surrounded by
big triangles are clusters with known optical substructures.}
\label{compare}
\end{figure}

\section{Cluster members selection and mass estimation}
To select the members of each systems and estimate the mass we use the
redshifts in the SDSS spectroscopic sample. The SDSS
spectroscopic pipeline (spectro1d) assigns a final redshift to each
object spectrum by choosing the emission or cross-correlation redshift
with the highest confidence level. The emission-redshift is obtained
matching the list of candidate emission lines against a list of common
galaxy and quasar emission lines. The cross-correlated redshift is
estimated by cross-correlating the spectrum with stellar, emission-line
galaxy, and quasar template spectra.

In order to select the cluster members, we proceed in two steps. In
the first step we follow the method of Girardi et al. (1993). Namely,
we select only galaxies within a circle of radius the Abell radius
(2.15 Mpc), and eliminate those with redshift $\mid cz -
cz_{cluster} \mid > 4000$ km s$^{-1}$, where $z_{cluster}$ is the mean
cluster redshift as given in the X-ray catalogues (see Section 2). We
then define the weighted gaps (see also Beers et al. 1990) in the
$z$-distribution of the remaining galaxies, and reject galaxies
separated from the main cluster body by a weighted gap $\geq 4$. This
allows us to define the cluster limits in velocity space.

In the second step of our procedure for membership selection, we
consider all galaxies (not only those within an Abell radius) with a
velocity within the limits defined with the gapper procedure, and
apply the method of Katgert et al. (2004) on these galaxies. The
method takes into account both the velocities and the clustercentric
positions of the galaxies (we take the X-ray center as the dynamical
center of the cluster). The method is identical to that of den Hartog
\& Katgert (1996) when the cluster sample contains at least 45 galaxies,
and it is a simplified version of it for smaller samples (for more
details, see Appendix A in Katgert et al. 2004).

The virial analysis (see, e.g., Girardi et al. 1998) is then performed
on the clusters with at least 10 member galaxies. The velocity
dispersion is computed using the biweight estimator (Beers et
al. 1990). The virial masses are corrected for the surface 
pressure term (The \& White 1986) by adopting a profile of Navarro et
al. (1996, 1997; NFW) with concentration parameter $c=4$ (this profile
has been found to describe the average mass profile of rich clusters
by Katgert et al. 2004). Correction for the surface term requires
knowledge of the $r_{200}$ radius, for which we adopt Carlberg et
al.'s (1997a) definition (see eq.(8) in that paper), as a first
guess. After the virial mass is corrected for the surface 
pressure term, we refine our $r_{200}$ estimate using the virial mass
density itself. Say $M_{vir}$ the virial mass (corrected for the
surface term) contained in a volume of radius equal to the
clustercentric distance of the most distant cluster member in the
sample, i.e. the aperture radius $r_{ap}$. The radius $r_{200}$ is
then given by:
\begin{equation}
r_{200} \equiv r_{ap} \, [\rho_{vir}/(200 \rho_c)]^{1/2.4}
\label{e-r200}
\end{equation}
where $\rho_{vir} \equiv 3 M_{vir}/(4 \pi r_{ap}^3)$ and $\rho_c(z)$
is the critical density at redshift $z$ in the adopted cosmology. The
exponent in eq.(\ref{e-r200}) is the one that describes the average
cluster mass density profile near $r_{200}$, as estimated by Katgert
et al. (2004) for an ensemble of 59 rich clusters. Similarly,
$r_{500}$ is estimated by setting 500 instead of 200 in
eq.(\ref{e-r200}).  Finally, a $c=4$ NFW profile is used to
interpolate (or, in a few cases, extrapolate) the virial mass
$M_{vir,c}$ from $r_{ap}$ to $r_{200}$ and $r_{500}$.
 
Our clusters span a wide range in mass; since clusters of
different masses have different concentrations (see, e.g. Dolag et al.
2004) we should in principle compute the cluster masses, $M$'s, using
a different concentration parameter $c$ for each cluster.  According
to Dolag et al. (2004), $c \propto M^{-0.102}$. Taking $c=4$ for
clusters as massive as those analysed by Katgert et al. (2004), $M
\simeq 2 \times 10^{15} M_{\odot}$, Dolag et al.'s scaling implies our
clusters span a range $c \simeq 3$--6. Using $c=6$ instead of $c=4$
makes the mass estimates 4\% and 10\% higher at, respectively,
$r_{200}$ and $r_{500}$, while using $c=3$ makes the mass estimates
lower by the same factors. This effect being clearly much smaller than
the observational uncertainties, we assume the same $c=4$ in the
analysis for all clusters.

Even if the completeness level of the SDSS spectroscopic sample is very high,
in the central regions of galaxy clusters such a level is
likely to drop because fibers cannot be placed closer than 55
arcsec. We estimate that the spectroscopic completeness drops to $\sim
70$\% in the central $\sim 0.1$ Mpc cluster regions. This affects the
observed number density profile of a cluster, and hence our estimate
of the projected mean harmonic pairwise separation $<R_{ij}^{-1}>$,
and, as a consequence, also our virial mass estimates (see, e.g.,
Beers et al. 1984). Using the average cluster number density profile,
and the relation between the core radius of this profile and
$<R_{ij}^{-1}>$, as given by Girardi et al. (1995, 1998), we estimate
that this effect of incompleteness translates into an average
over-estimate of the virial mass of only $\sim 5$\%. Since the effect
is very small, and much smaller than the observational uncertainties,
we neglect this correcting factor in the following analysis.
 
There are several indications that cluster galaxies of later
morphological and/or spectral type, have wider velocity distributions
than early-type cluster galaxies (Moss \& Dickens 1977; Biviano et al.
2002 and references therein). As a consequence, we also estimate the
virial masses by considering only cluster members along the red
sequence in the $u-i$ vs. $i$ colour-magnitude diagram.  When only the
red-sequence cluster members are used to compute the virial masses,
these masses are $\sim 25$\% smaller than when all the cluster members
are used. A similar effect has been discussed by Biviano et al. (1997)
for the galaxy clusters of the {\em ESO Nearby Abell Cluster Survey,}
and by Carlberg et al. (1997b) for the galaxy clusters of the CNOC.
The effect is generally interpreted as evidence for ongoing accretion
of blue field galaxies onto the cluster, before complete virialization
(see, e.g., the discussion in Biviano \& Katgert 2003).
Since this effect is not negligible, in the following we consider both
the mass estimates obtained using only the red-sequence galaxies, and
the mass estimates obtained using all the cluster members. However,
for the sake of conciseness, we only plot results for the mass estimates
obtained using all the cluster members.

\begin{figure}
\begin{center}
\begin{minipage}{0.5\textwidth}
\resizebox{\hsize}{!}{\includegraphics{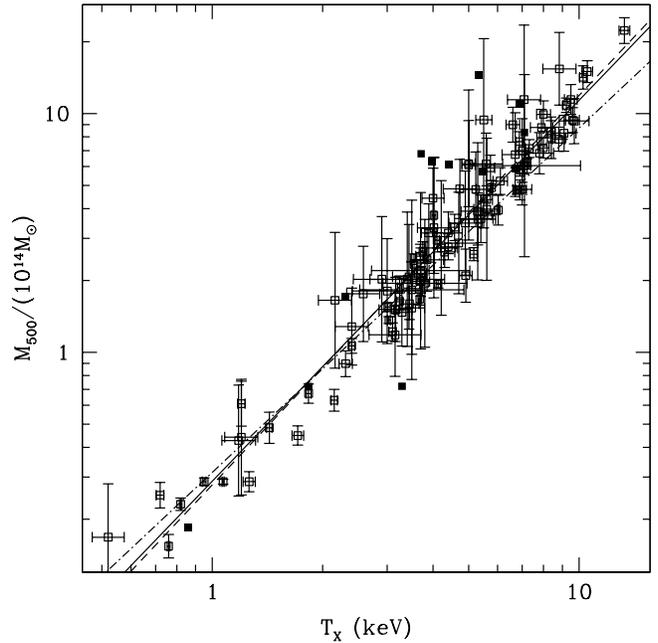}}
\end{minipage}
\end{center}
\caption{
Mass-Temperature relation calibrated with the cluster mass and
temperature used in R02. The new normalization in the $M-T_X$ relation is
60\% higher than in the classical $M-T_X$ relation of F01. The solid line
is the best fit line obtained using the enlarged sample of R02. The
dotted line is the best fit line obtained fitting the clusters with
$T_X$ higher than 4.5 keV, while the dashed-dotted line is the best
fit line derived from the fit of systems with temperature lower than
4.5 keV.}
\label{M_T}
\end{figure}

To check the consistency of our mass estimates with those obtained
from X-ray data, we retrieve the latter from R02, for a subsample of
10 clusters with optical mass estimates.  Fig. \ref{compare} shows the
overall agreement between the X-ray mass from R02 and the optical mass
calculated in this work (empty squares). The figure shows also
the same comparison between optical mass and the mass obtained from
the $M-T$ relation for 6 more clusters with know ASCA temperature but
with unknown X-ray mass (filled squares).  It is mainly for the
systems with evident substructures in the redshift distribution
(the empty and filled squares surrounded by big triangles in the
Figure) that the two mass estimates are in disagreement, as
expected. In the following analysis we do not consider these clusters
with strong optical substructures. We omit in fig.\ref{compare}
the comparison between the mass estimated from the $M-T_X$ relation
(see eq. \ref{eq1} below) and the direct measure of the X-ray mass
(R02). They are in very good agreement since the $M-T_X$ relation is
derived from R02 data, and the scatter in the relation is very small
as showed by the errors in eq. \ref{eq1} and the fig.\ref{M_T}.
 
\begin{figure}
\begin{center}
\begin{minipage}{0.5\textwidth}
\resizebox{\hsize}{!}{\includegraphics{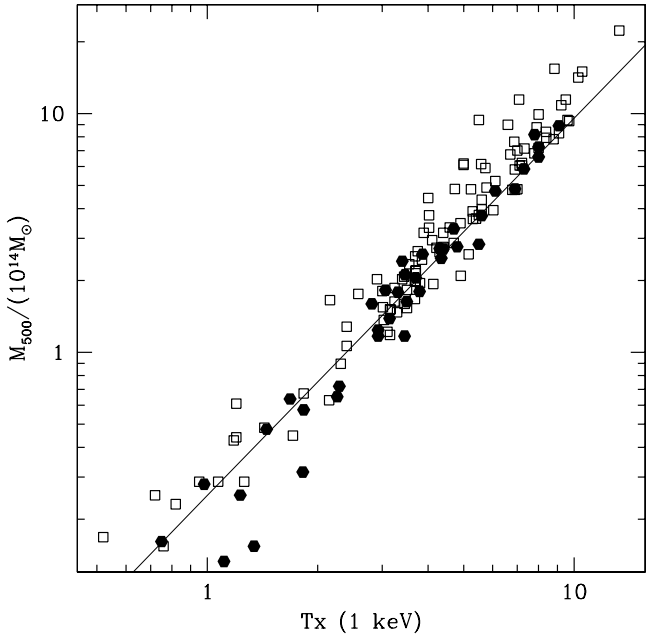}}
\end{minipage}
\end{center}
\caption{  
Relation between mass and temperature.  The mass is calculated using a
$\beta$-model and the temperature of the ICM. The empty squares are
the clusters of R02, while the filled hexagons are taken from F01. The
solid line in the figure is the best fit line given by F01.}
\label{R02_F01}
\end{figure}

\begin{figure}
\begin{center}
\begin{minipage}{0.5\textwidth}
\resizebox{\hsize}{!}{\includegraphics{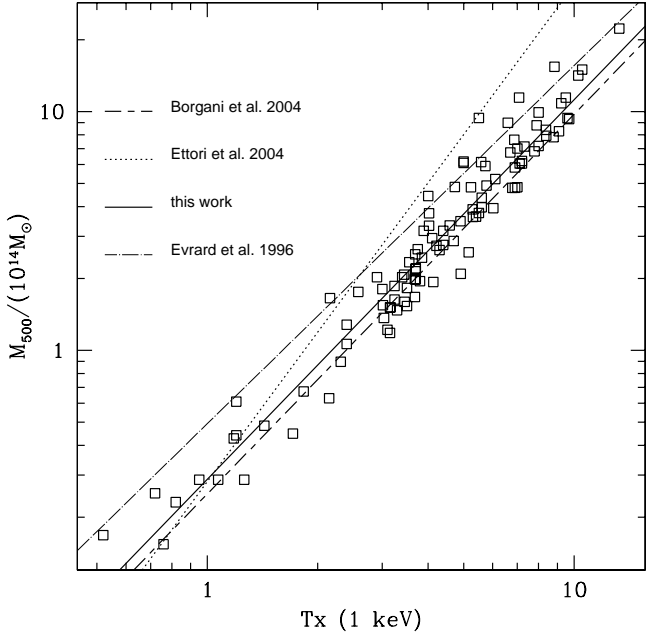}}
\end{minipage}
\end{center}
\caption{
Comparison of  the $M-T_X$ relation obtained with  the sample from R02 and
the  theoretical  prediction  from  Borgani  et al.  (2004),  Ettori  et
al. (2004), Evrard et al. (1996).}
\label{R02_teo}
\end{figure}

\subsection{Masses estimated from the $M-T_X$ relation}

We use the $M-T_X$ relation to estimate the mass for the subsample of
49 clusters with known ASCA temperature. First, we consider the
$M-T_X$ relations provided by Finoguenov et al. (2001, hereafter F01).
F01 provided several $M-T_X$ relations estimated using different
samples of systems: the cluster sample of R02 (HIFLUGCS + 60 more
clusters) and a sample of 39 clusters with known temperature profile
from ASCA data (Markevitch et al. 1998). We notice that the
masses estimated with the $M-T_X$ relation of F01 obtained from
clusters with known temperature profile are systematically smaller by
a factor 1.5-2 than both the virial masses obtained from the analysis
of the galaxy distribution, and also the X-ray masses of R02. F01's
sample comprises only 9 of the 45 clusters included in R02's sample
with temperature higher than 5 keV. The masses of these 9 clusters are
in good agreement with the masses estimated in R02 (which assumes the
isothermality of the ICM) but, as shown in Fig. \ref{R02_F01}, the
high-mass region is not well sampled.  Moreover, the presence of 4
systems in the low mass regime with $\beta \sim 0.3$ (where
$\beta$ is the exponent of the $\beta$-model used to calculate the
X-ray mass, see F01 or R02 for more details) implies both
a steepening of the $M-T_X$ relation and a decrease of its
normalization.

Therefore, instead of using the $M-T_X$ relation of F01, we prefer to
recalibrate the $M-T_X$ relation using the data of R02 in order to
obtain both the $M_{500}-T_X$ and the $M_{200}-T_X$ relations. Data
are fitted with the ODRPACK routine (Akritas $\&$ Bershady
1996) and the errors are calculated using a bootstrap method. In the
following analysis the mass is calculated from the temperature with
the following relations:
\begin{equation}
M_{500}=2.89\pm 0.29 \times 10^{13} T_X^{1.59\pm 0.04} 
\label{eq1}
\end{equation}
\begin{equation}
M_{200}=4.69\pm 0.36 \times 10^{13} T_X^{1.59\pm 0.05} 
\end{equation}
The normalization of the $M-T_X$ relation in eq.(\ref{eq1}) is 60\%
higher than that of the usual $M-T_X$ relation of F01 estimated with
the sample with known $T_X$ profiles. There is much better agreement
(within 1 $\sigma$) with the relation of F01 estimated excluding from
the sample the groups with $\beta$ lower than 0.4. Fig. \ref{M_T}
shows the best fit for the enlarged sample of R02 while
Fig. \ref{R02_teo} shows the comparison of our best fit with the
relation predicted by the hydrodynamical simulations. 

\begin{table*}
\begin{center}
\begin{tabular}[b]{cc|c|ccccc|ccccc}
\hline
\multicolumn{2}{c|}{A-B relation}& \multicolumn{1}{c|}{sample}& \multicolumn{5}{c|}{red members}& \multicolumn{5}{|c}{all members} \\ \hline
\renewcommand{\arraystretch}{0.2}\renewcommand{\tabcolsep}{0.05cm}
A& B& &$\alpha $ & $\beta$ & $\sigma$ &$\sigma _B$ & $\sigma_A$ & $\alpha $ & $\beta$ &$\sigma$& $\sigma _B$ & $\sigma_A$\\
\hline
$M_{500}$&$L_{op}$& O &$ 0.80\pm 0.04$ & $-0.03 \pm 0.03$ & 0.12 & 0.16 & 0.16 &$0.81 \pm 0.04$ & $-0.09 \pm  0.03$ & 0.10 & 0.16 & 0.15 \\
 & &                X &$ 0.96\pm 0.05$ & $0.01  \pm 0.04$ & 0.09 & 0.15 & 0.16 \\
 & &                E &$ 0.91\pm 0.04$ & $-0.03 \pm 0.03$ & 0.11 & 0.19 & 0.17 &$0.90 \pm 0.05$ & $-0.06 \pm  0.03$ & 0.11 & 0.17 & 0.16 \\
\hline	
$M_{200}$&$L_{op}$& O &$ 0.79\pm 0.04$ & $0.12 \pm 0.03$ & 0.13 & 0.16 & 0.15 & $0.79 \pm 0.04$ & $-0.05 \pm  0.04$ & 0.10 & 0.16 & 0.14 \\
 & &                X &$ 1.05\pm 0.07$ & $-0.04 \pm 0.06$ & 0.10 & 0.17 & 0.16 \\
 & &                E &$ 0.91\pm 0.04$ & $0.05 \pm 0.04$ & 0.11 & 0.17 & 0.16 & $ 0.91\pm 0.04$ & $-0.08 \pm 0.04$ & 0.12 & 0.19 & 0.17\\
\hline
$M_{500}$&$L_X   $& O &$ 1.30\pm 0.09$ & $-0.77 \pm 0.09$ & 0.15 & 0.19 & 0.30 & $1.41 \pm 0.12$ & $-0.92 \pm  0.06$ & 0.18 & 0.19 & 0.32 \\
 & &                X &$ 1.87\pm 0.12$ & $-0.83  \pm 0.06$ & 0.14 & 0.19 & 0.29 \\
 & &                E &$ 1.68\pm 0.09$ & $-0.88 \pm 0.06$ & 0.17 & 0.22 & 0.36 &$1.69 \pm 0.10$ & $-1.00 \pm  0.06$ & 0.19 & 0.23 & 0.43 \\
& &                 R+E & $ 1.50\pm 0.05$ & $-0.38  \pm 0.03$ & 0.17 & 0.22 & 0.27 \\
\hline
$M_{200}$&$L_X$   & O &$ 1.30\pm 0.09$ & $-0.95 \pm 0.07$ & 0.15 & 0.19 & 0.30 & $1.32 \pm 0.08$ & $-1.08 \pm  0.07$ & 0.17 & 0.18 & 0.30 \\
 & &                X &$ 1.98 \pm 0.13$ & $-1.24 \pm 0.10$ & 0.16 & 0.20 & 0.29 \\
 & &                E &$ 1.71 \pm 0.09$ & $-1.18 \pm 0.07$ & 0.17 & 0.21 & 0.35 & $ 1.66\pm 0.09$ & $-1.27 \pm 0.07$ & 0.17 & 0.21 & 0.39\\
& &                 R+E & $ 1.58\pm 0.07$ & $-0.92  \pm 0.06$ & 0.16 & 0.26 & 0.30 \\
\hline
$L_X $&$L_{op}$    & O &$ 0.60 \pm 0.05$ & $0.40 \pm 0.02$ & 0.12 & 0.30 & 0.16 & $0.63 \pm 0.04$ & $0.41 \pm  0.02$ & 0.14 & 0.29 & 0.17 \\
 &$(r_{500})$ &                X &$ 0.53 \pm 0.03$ & $0.39 \pm 0.03$ & 0.16 & 0.33 & 0.19 \\
 & &                E &$ 0.54 \pm 0.03$ & $0.40 \pm 0.03$ & 0.14 & 0.33 & 0.18 & $ 0.55\pm 0.03$ & $0.41 \pm 0.02$ & 0.15 & 0.33 & 0.18\\
\hline
$L_X$&$L_{op}$& O &$ 0.59 \pm 0.05$ & $0.55 \pm 0.02$ & 0.13 & 0.28 & 0.16 & $0.64 \pm 0.04$ & $0.57 \pm  0.02$ & 0.15 & 0.27 & 0.17 \\
 &$( r_{200})$ &                X &$ 0.55 \pm 0.04$ & $0.56 \pm 0.03$ & 0.19 & 0.35 & 0.22 \\
 & &                E &$ 0.56 \pm 0.03$ & $0.56 \pm 0.03$ & 0.16 & 0.32 & 0.18 & $ 0.58\pm 0.03$ & $0.57 \pm 0.02$ & 0.17 & 0.31 & 0.19\\
\hline
$r_{500}$&$L_{op}$& O& $2.28\pm0.14$ & $0.31\pm 0.02$ &  0.05& 0.05&  0.15 & $2.26\pm0.13$ & $0.33\pm 0.04$ &  0.05& 0.06&  0.14 \\ 
 & &                X& $2.95\pm0.15$ & $0.43\pm 0.02$ &  0.04& 0.05&  0.14 \\
 & &                E& $2.53\pm0.12$ & $0.36\pm 0.02$ &  0.05& 0.07&  0.17 & $2.50\pm0.13$ & $0.38\pm 0.05$ &  0.05& 0.06&  0.17 \\ 
\hline
$r_{200}$&$L_{op}$& O&$2.25\pm0.13$ & $0.07\pm 0.04$ &  0.06& 0.07&  0.14 &$2.28\pm0.14$ & $0.08\pm 0.04$ &  0.06& 0.08&  0.14\\ 
 & &                X&$2.88\pm0.18$ & $0.06\pm 0.05$ &  0.06& 0.08&  0.17 \\ 	
 & &                E&$2.49\pm0.12$ & $0.06\pm 0.04$ &  0.07& 0.09&  0.17 &$2.52\pm0.11$ & $0.07\pm 0.04$ &  0.07& 0.08&  0.17 \\ 
\hline
\end{tabular}
\caption{
The table lists the best fit parameters for the $L_{op}-M$, $L_X-M$,
$L_{op}-L_X$ and $L_{op}-r_{500/200}$ relations respectively for
different samples of galaxy clusters and for different methods.  The O
letter refers to the sample with masses estimated from the dynamical
analysis performed with the optical spectroscopic data.  The letter X
refers to the sample with masses estimated from the $M-T_X$
relation. The letter E refers to the enlarged sample, which comprises
all the clusters in the RASS-SDSS galaxy cluster catalog with known
mass.  The $R+E$ refers to the enlarged sample plus the sample of R02.
The left side of the table lists the results obtained performing the
dynamical analysis only on the red members od the system.  The right
side lists the best fit values of the correlations obtained using the
results of the dynamical analysis applied to all the cluster
members. The table lists three estimations of the scatter for each
relation: $\sigma$ is the orthogonal scatter of the A-B relation,
$\sigma_A$ is the scatter in the A variable and $\sigma_B$ is the
scatter in the B variable. All the scatters in the table are
expressed in dex, while all the errors are given at the 95\%
confidence level.}
\label{table1}
\end{center}							   
\end{table*}	

\section{Correlation of the optical and X-ray luminosities with the cluster global properties}
In this section we examine the correlation of the optical luminosity
$L_{op}$ and the X-ray luminosity $L_X$ with quantities derived from
the optical and X-ray data, such as the total mass, the velocity
dispersion, the X-ray temperature, $r_{500}$, and $r_{200}$. The main
motivation in deriving these dependences is to use $L_{op}$ and $L_X$,
as predictors of the other quantities. Moreover, we will compare the
quality of the two quantities $L_{op}$ and $L_X$ as predictors.  To
quantify all the dependences, a linear regression in log-log space is
performed using a numerical orthogonal distance regression method
(ODRPACK). The fits are performed using the form
\begin{equation}
\log(L_{op}/(10^{12}L_{\odot}))=\alpha \times \log(P_C)+{\beta}
\end{equation}
\begin{equation}
\log(L_{X}/(10^{44}erg s^{-1}))=\alpha \times \log(P_C)+{\beta}
\end{equation}
where $P_C$ is the cluster global property, and the errors of each
variable are transformed into log space as $\Delta
\log(x)=\log(e)(x^+-x^-)/(2x)$, where $x^+$ and $x^-$ denote the upper
and lower boundary of the error range of the quantity, respectively.
In all the correlations we analyse separately the sample with masses
derived from the dynamical analysis (69 clusters) and the sample with
masses derived from the $M-T_X$ relation (49 systems). Then we analyse
the correlations obtained using all the clusters of the RASS-SDSS
galaxy cluster sample with known mass (either the virial estimate from
optical data, or, when this is not available, the mass derived from
the X-ray temperature) for a total number of 102 systems (69
cluster with known optical mass $+$ 49 clusters with mass estimated
from the temperature $-$ 16 common clusters) . In the following
tables, 'O' (Optical sample) refers to the sample with optical masses,
the letter 'X' (X-ray sample) to the sample with mass derived from
$T_X$ and, finally, the letter 'E' (Enlarged sample) refers to the
sample of all the clusters with known mass.

In all the following analysis we consider only the relation obtained
with the optical luminosity calculated in the $i$ Sloan band. All the
results obtained in the other three SDSS optical bands ($g, r$ and $z$)
are listed in the Appendix.

\begin{figure}
\begin{center}
\begin{minipage}{0.5\textwidth}
\resizebox{\hsize}{!}{\includegraphics{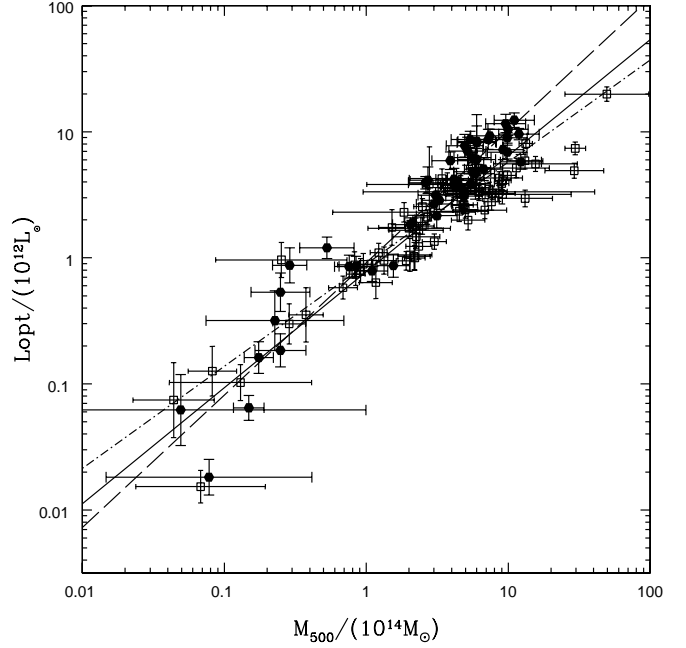}}
\end{minipage}
\end{center}
\caption{ $L_{op}-M_{500}$ relation in the $i$ Sloan band.  The empty
squares in the figure indicate clusters with mass estimated from the
dynamical analysis of the Sloan spectroscopic data.  The filled points
indicate systems with mass estimated from the $M-T_X$ relation. The
dot-dashed line is the best fit line obtained for the O sample. The
dashed line is the best fit line for the X sample, while the solid
line is the result obtained from the E sample. }
\label{M_LO_spi}
\end{figure}

\begin{figure}
\begin{center}
\begin{minipage}{0.5\textwidth}
\resizebox{\hsize}{!}{\includegraphics{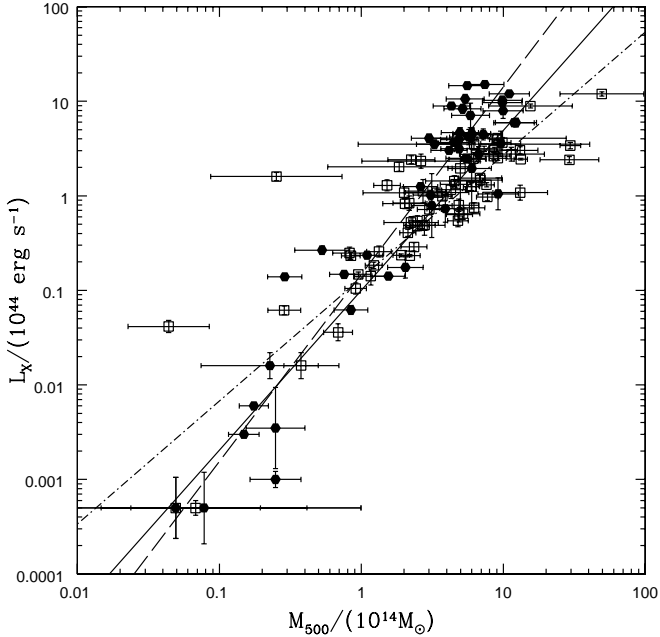}}
\end{minipage}
\end{center}
\caption{ $L_X-M_{500}$ relation for the RASS-SDSS galaxy cluster
sample.  The empty squares in the figure are clusters with mass
estimated from the dynamical analysis of the Sloan spectroscopic data.
The filled points are systems with mass estimated from the $M-T_X$
relation. The dot-dashed line is the best fit line obtained for the O
sample. The dashed line is the best fit line for the X sample, while
the solid line is the result obtained from the E sample.}
\label{LX_M}
\end{figure}

\begin{figure}
\begin{center}
\begin{minipage}{0.5\textwidth}
\resizebox{\hsize}{!}{\includegraphics{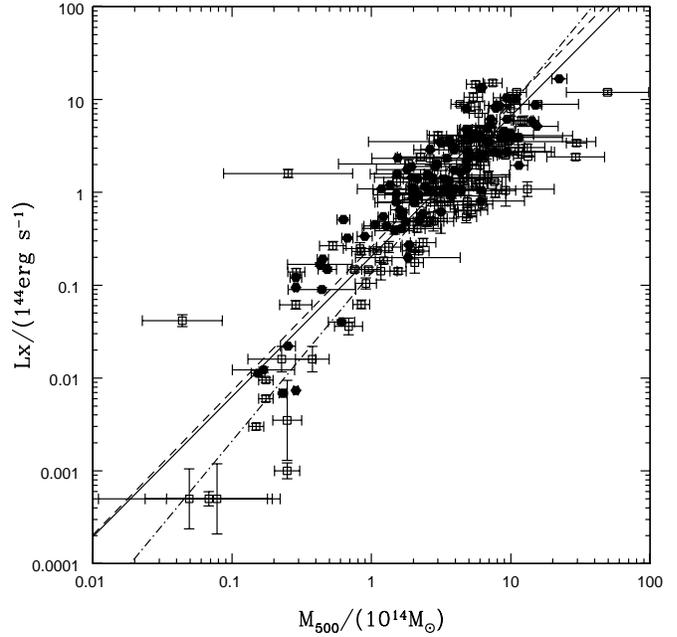}}
\end{minipage}
\end{center}
\caption{  
$L_X-M_{500}$ relation for the RASS-SDSS galaxy cluster sample plus
the clusters sample of R02.  The empty squares are
the E clusters, while the filled hexagons are the clusters of the R02.
The dot-dashed line is the best fit line obtained for the E
sample. The dashed line is the best fit line for the R02 sample, while
the solid line is the result obtained from the 'E+R' sample.  }
\label{M_LX}
\end{figure}

\begin{figure}
\begin{center}
\begin{minipage}{0.5\textwidth}
\resizebox{\hsize}{!}{\includegraphics{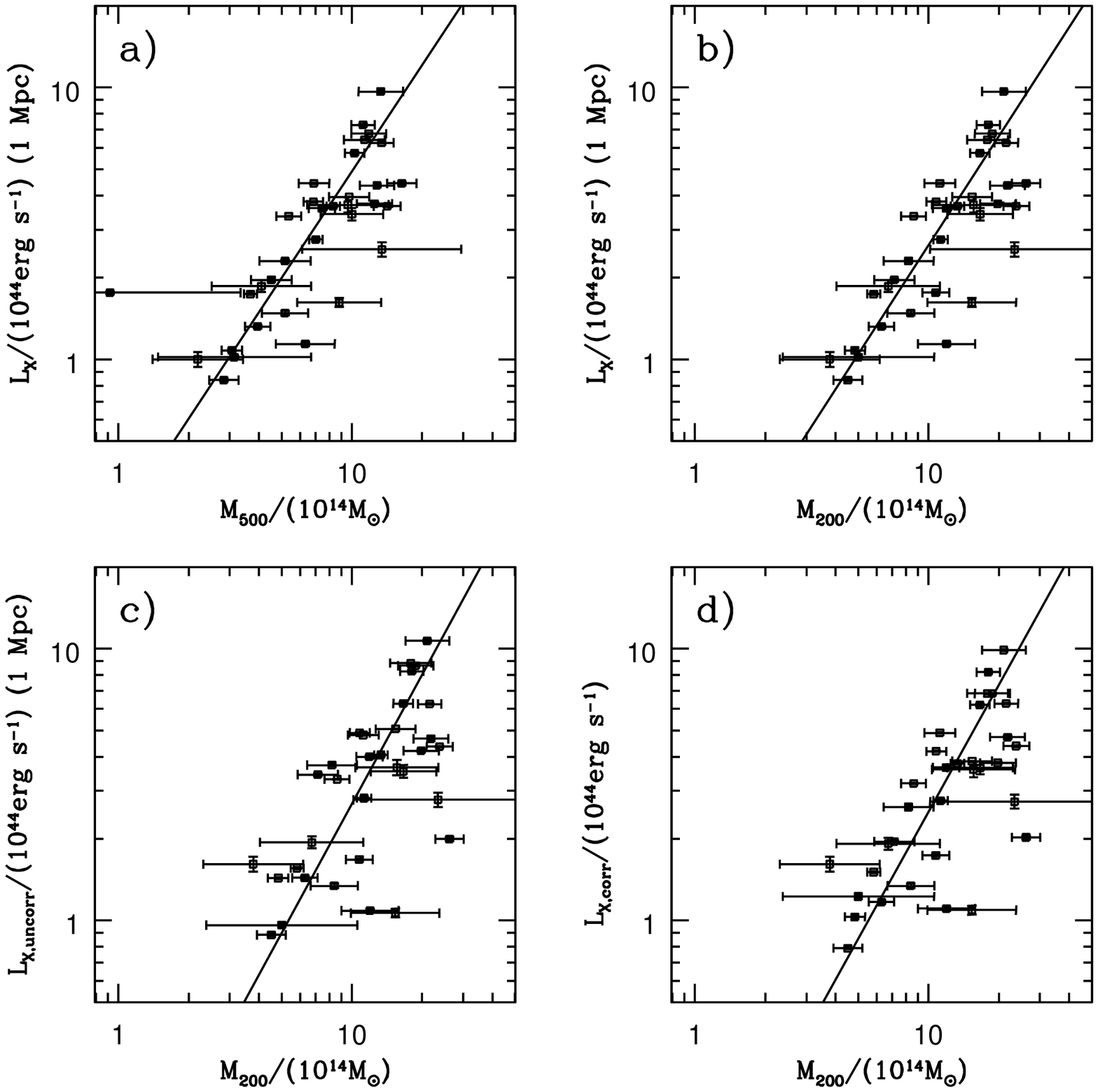}}
\end{minipage}
\end{center}
\caption{  
$L_X-M_{500}$ relation for the sample of M98. The
$L_X-M$ relation is obtained using a cool-core-corrected X-ray
luminosity.  Panels $a$ and $b$ show the $L_X-M$ relation given
by the X-ray luminosity calculated in the $0.1-2.4$ keV energy band
within 1.4 Mpc from the cluster center with $M_{500}$ and $M_{200}$
respectively.  The X-ray luminosity is taken from M98,
while the mass is taken from R02. R02 used the cool-core-corrected
temperature of M98 to calculate the mass.  Panels $c$ and
$d$ show the same relation given by the total X-ray luminosity of the
system (calculated within the physical size of the system) in the
$0.1-2.4$ keV energy band and $M_{200}$. The total $L_X$ is taken from
R02 and is calculated with the same method used to estimate the X-ray
luminosity in the RASS-SDSS sample.  Panel $c$ shows the $L_X-M$
relation with uncorrected X-ray luminosity, while panel $d$ shows the
$L_X-M$ relation obtained using the X-ray luminosity corrected for the
cool-core effect.  The correction is obtained comparing the
corrected and uncorrected $L_X$ retrieved in M98.}
\label{Mark}
\end{figure}

\subsection{Correlations of the optical and X-ray luminosities with the cluster mass.}
Both the optical and the X-ray luminosity show a tight relation with
the cluster mass (see Figs. \ref{M_LO_spi} and \ref{LX_M},
respectively for the $L_{op}-M_{500}$ and $L_X-M_{500}$
relations). The optical luminosity is estimated within the same radius
of the mass, while $L_X$ is the {\em total} X-ray luminosity of the
system. The total $L_X$ is not estimated within a fixed aperture but
is calculated from the X-ray luminosity radial profile. Table
\ref{table1} lists the best fit values of the correlation for the
different samples. In both $L_{op}-M$ and $L_X-M$ relations the slope
is flatter for the O sample than for the X sample.  As a consequence,
the best fit values of the correlation obtained with the enlarged
sample are a mean of the previous values. However, all the derived
values of slope and normalization are in agreement within 1.5
$\sigma$.  The slope of $0.75\pm 0.02$ of the $L_{op}-M$ relation in
the $B$ band studied by Girardi et al. (2002) is in very good
agreement (within 1 $\sigma$) with the slope of $0.81\pm0.04$ obtained
with the O sample. The agreement is less good (within 3 $\sigma$) for
the slope obtained with the X and E samples.  The values of slope and
normalization of the $L_X-M$ relation lies in the range of values
given in R02. To recalibrate the $L_X-M$ relation we add our enlarged
sample to the sample of clusters of R02 obtaining a final sample of
198 clusters (106 from R02 $+$ 102 from this work $-$ 10 common
systems) with known mass and X-ray luminosity. The best fit values
of the correlation obtained with this sample are indicated by 'R+E' in
Table \ref{table1}. The resulting slope of $1.5\pm 0.05$ is in good
agreement with the value obtained from the E sample. The correlation
is shown in Fig. \ref{M_LX}.

\begin{figure}
\begin{center}
\begin{minipage}{0.5\textwidth}
\resizebox{\hsize}{!}{\includegraphics{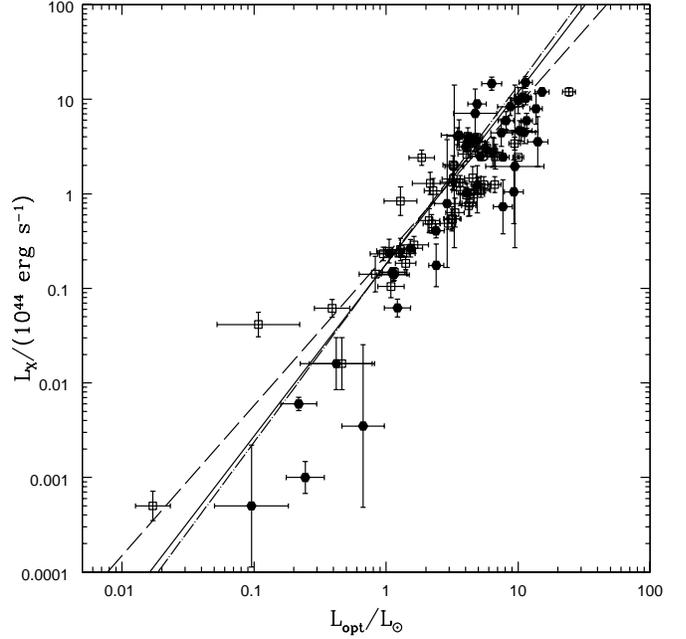}}
\end{minipage}
\end{center}
\caption{  
$L_{op}-L_X$ relation.  The optical luminosities is calculated within
$r_{500}$. The empty squares in the figure are clusters with mass
estimated from the dynamical analysis of the Sloan spectroscopic
data. The filled points are systems with
$r_{500}$ estimated from mass obtained with the $M-T_X$ relation.  The
dot-dashed line is the best fit line obtained for the O sample.  The
dashed line is the best fit line for the X sample, while the solid
line is the result obtained from the E sample.}
\label{LX_LO}
\end{figure}

It is particularly interesting to compare the scatter of the
$L_{op}-M$ and $L_X-M$ relations in order to understand which of the
two observables, $L_{op}$ and $L_X$, shows the best correlation with
the cluster mass. Table \ref{table1} lists three kinds of scatters
evaluated for each correlation: the orthogonal scatter, which gives an
estimate of the dispersion along the direction orthogonal to the best
fit line, and the scatter in both variables. The orthogonal scatter in
the $L_{op}-M$ relation has a minimum value of 20\% and a maximum of
30\% in all the analysed correlations. The observable $L_{op}$ can be
used to predict the cluster mass within $r_{500}$ or $r_{200}$ with a
maximum accuracy of 40\% and with a minimum of 50\%. The X-ray
luminosity shows a less tight relation with the cluster mass for all
the analysed correlations. In fact, the orthogonal scatter of the
$L_X-M$ relations lies in the range 38-50\%, while the mass can be
predicted from the X-ray luminosity with an accuracy in the range
55-65\%. These values are in agreement with the results of R02 and are
confirmed also by the correlation obtained with the 'R+E' sample, as
indicated in Table \ref{table1}. The dispersion along the $L_X$ axis
is much larger (more than 90\%) and is due to the propagation of the
errors, since $\sigma_{L_X} \sim \alpha \times \sigma_{M}$. The
difference in the scatter observed in the two relations cannot be
explained by the measurement errors, since the average error in the
$L_{op}$ measure is around $15\%$ and is comparable to the average
measurement error in $L_X$ ( around $10\%$).

\subsection{The scatter in  the $L_{op}-M$ and $L_X-M$ relations}
A possible explanation for the large scatter of the $L_X-M$ relation
is the presence of a large number of groups in our samples.  In fact
measuring the mass and the X-ray luminosity for low-mass systems is
not an easy task.  However, the low-mass systems do not increase the
scatter in the $L_{op}-M$ relation, as they should if their masses
were not measured correctly.  Moreover, the scatter of the $L_X-M$
relation is the same in all the mass ranges, since including or
excluding the groups in our analysis does not change the amount of
scatter in the correlations.  A more plausible explanation for the
larger scatter in the $L_X-M$ relation in comparison to the $L_{op}-M$
relation, is the presence of a large number of cool-core (once named
'cooling flow') clusters in the analysed sample.  In fact, the
presence of a cool core in a cluster could cause an increase of the
observed X-ray luminosity for a given cluster mass.  In principle even
the mass estimated from the ICM temperature could be affected by this
effect.  In fact, Markevitch (1998, hereafter M98) has shown that the
presence of a cool core in a system can significantly affect the
temperature estimation if the cool core region is not
excised. However, the scatter observed in the $L_X-M$ relation
obtained with the X sample and with the R02 sample, in which the mass
is calculated using the temperature, is exactly the same observed in
the $L_X-M$ relation obtained with the O sample, in which the mass
estimation is not influenced by the presence of a cool core. Thus, we
do not expect that the presence of a cool core affects the mass in our
correlation.  Therefore, in Fig. \ref{M_LX} the cool-core clusters
should move to higher X-ray luminosity but not to higher mass.  We
call this effect the 'cool core' effect throughout the paper.

\begin{table*}
\begin{center}
\begin{tabular}[b]{c|ccccc}
\hline
\renewcommand{\arraystretch}{0.2}\renewcommand{\tabcolsep}{0.05cm}
&$\alpha $ & $\beta$ & $\sigma$ &$\sigma _B$ & $\sigma_A$ \\
\hline
$L_X(0.1-2.4 keV)-M_{500}$ (1.4 Mpc) &$ 1.30\pm 0.12$ & $-0.61 \pm 0.10$ & 0.10 & 0.11 & 0.20 \\
\hline
$L_X(0.1-2.4 keV)-M_{200}$ (1.4 Mpc) &$ 1.33\pm 0.13$ & $-0.91 \pm 0.14$ & 0.10 & 0.11 & 0.14 \\
\hline
$L_X(Bol)-M_{200}$ (1.4 Mpc) & $ 2.01\pm 0.20$ & $-1.35 \pm 0.22$ & 0.10 & 0.11 & 0.20 \\
\hline	
$L_{X,corr}(0.1-2.4 keV)-M_{200}$ (tot) & $ 1.55\pm 0.19$ & $-1.15 \pm 0.20$ & 0.11 & 0.12 & 0.17 \\
\hline	
$L_{X,uncorr}(0.1-2.4 keV)-M_{200}$ (tot) & $ 1.58\pm0.23 $ & $-1.15\pm0.24$ & 0.11 & 0.13 & 0.18 \\
\hline
\end{tabular}	
\caption{ 
The table lists the best fit values for the $L_X-M$ relations.  The
first three lines list the relations obtained with M98's data:
$L_X-M_{500}$ with $L_X$ calculated in 0.1-2.4 keV energy band, within
1.4 Mpc from the cluster center and corrected for cool core effect,
$L_X-M_{200}$ ($L_X$ before), $L_X(Bol)-M_{200}$ with the bolometric
X-ray luminosity calculated within 1.4 Mpc from the cluster center and
corrected for cool core effect. The last two lines list the
correlations obtained with the total X-ray luminosity taken from R02
corrected and uncorrected for cool core effect respectively. The
table lists three estimation of the scatter for each relation:
$\sigma$ is the orthogonal scatter of the A-B relation, $\sigma_A$ is
the scatter in the A variable and $\sigma_B$ is the scatter in the B
variable. All the scatters in the table are expressed in dex, while
all the errors are given at the 95\% confidence level. }
\label{table3}
\end{center}							   
\end{table*}

Unfortunately our data are not able to fully explore this
effect.  To calculate the amount of scatter due to the effect of cool
core on the X-ray luminosity, we must use the cluster sample of M98.  In
this sample the X-ray luminosities and temperatures are corrected for
the cool core effect.  We retrieve the masses of 33 of the 35 clusters
of that sample from R02.  The masses taken from R02 are all calculated
with the corrected temperature of M98.  In Fig.  \ref{Mark} we show
the $L_X-M$ relation obtained using a cool-core-corrected X-ray
luminosity. Panels $a$ and $b$ show the $L_X-M$ relation given by the
X-ray luminosity calculated in the $0.1-2.4$ keV energy band within
1.4 Mpc from the cluster center with $M_{500}$ and $M_{200}$
respectively.  Panels $c$ and $d$ show the same relation given by the
total X-ray luminosity of the system in the same energy band and
$M_{200}$. The total $L_X$ is taken from R02 and is calculated within
$r_{200}$ with a method similar to the method used to estimate the
X-ray luminosity in the RASS-SDSS sample. Therefore, it should give a
robust estimate of the total X-ray emission of the system.  Panel $c$
shows the $L_X-M$ relation with uncorrected X-ray luminosity, while
panel $d$ shows the $L_X-M$ relation obtained using the X-ray
luminosity corrected for the cool core effect.  The correction is
obtained by comparing the corrected and uncorrected $L_X$ retrieved in
M98. M98 removed the effect due to the presence of a cool
core with the excision of the cool core region.  They assumed that a
70 kpc radius contains most of the cool core emission in non-extreme
cool core clusters, such as those in the sample considered here.
Therefore, to do the excision in a uniform manner, for all clusters,
regions of 70 kpc radius centered on the main brightness peak were
masked, and the resulting fluxes and luminosities were multiplied by
1.06 to account for the flux inside the masked region, assuming an
average $\beta$ model for the cluster X-ray brightness.  Therefore,
subtracting the corrected X-ray luminosity from the uncorrected one
gives the amount of cool-core correction applied by M98. We use the
same amount to correct the total X-ray luminosities given by R02.

As shown in Table \ref{table3}, applying the correction for cool-core
effect does not change the scatter of the relation. In fact, in the
relation obtained using the total X-ray luminosity, the scatter along
the $M_{200}$ axis is 0.13 dex before the correction and 0.12 dex
after that.  This means that the cool core correction can reduce the
scatter by only 3\%. As a matter of fact, for most of the clusters in
the M98 sample the correction is of the same order.  Moreover, it is
important to stress here that even the scatter of the $L_X-M$ relation
obtained in this analysis with the uncorrected X-ray luminosity is
much lower than the dispersion obtained with the the RASS-SDSS galaxy
cluster catalog, which is a sample more than 3 times larger. In fact,
the sample of M98 covers a very small range in mass and X-ray
luminosity, only one order of magnitude in both variables. Hence, the
statistical significance of the result is very low and it cannot be
taken as a robust result, since it does not seem to represent the
behavior of the $L_X-M$ relation obtained with much larger samples of
clusters.  Therefore, it is not clear if the presence of a large
number of cool core clusters in our sample and in the sample of R02
could really contribute to the scatter in the $L_X-M$ relation. 
As a last point, we notice that replacing the $L_X$ calculated in the
ROSAT energy band (0.1-2.4 keV) with the bolometric luminosity does
not change at all the scatter in the relation. The bolometric X-ray
luminosity is taken from M98. The slope of the $L_{X,bol}-M_{200}$
relation is steeper then the $L_{X,ROSAT}-M_{200}$ relation as
expected (the bolometric correction is smaller for the faint X-ray
clusters than in the bright ones), while the orthogonal scatter and
the dispersion along the $M_{200}$ axis are unchanged. The scatter in
the $L_X$ variable changes because of the slope, since $\sigma_B \sim
\alpha \times \sigma_A$. In conclusion, to really understand the
nature of the scatter in the $L_X-M$ relation and its connection to
the cool core correction to the X-ray luminosity, the analysis should
be done with a cluster sample much larger than the M98 sample and with
a much more extended range in mass and $L_X$.

Finally, let us consider the possibility that the better behavior of
the optical luminosity as a mass predictor, in comparison with $L_X$,
could be due to the fact that $L_{op}$ is calculated within the same
aperture of the mass ($r_{500}$ and $r_{200}$) while $L_X$ is
estimated within a variable aperture. R02 calculate the mass within
$r_{200}$ and yet obtain the same scatter we observe in the $L_X-M$
relation. Therefore, the scatter in the $L_X-M$ relation does not seem
to depend on the limiting radius used to compute $L_X$. The
observed dispersion in the $L_X - M$ relation is most probably due to
variations in the compactness of clusters. The dichotomy of compact cD
clusters (often associated with the cooling flow signatures) and less
compact non-cD clusters is more pronounced and significant than just
an excess of X-ray flux in the central 70 kpc region (as used in the
above correction for cooling flows). This is indicated for example by
the work of Jones \& Forman (1984) and Ota \& Mitsuda (2002) and
discussed by Fabian et al. (1994). Due to the strong quadratic
dependence of the X-ray emission on the gas density this variation in
compactness is observed in an amplified way in the X-ray luminosity
variation.

\begin{figure}
\begin{center}
\begin{minipage}{0.5\textwidth}
\resizebox{\hsize}{!}{\includegraphics{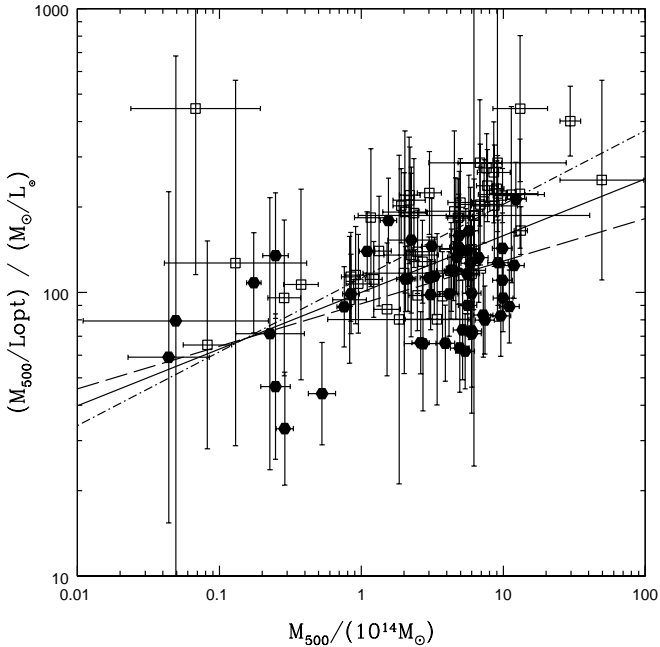}}
\end{minipage}
\end{center}
\caption{  
$M/L-M$ relation. The mass-to-light ratio is calculated the $i$ Sloan
band.  Mass and optical luminosities are
calculated within $r_{500}$.  The empty squares in the figure are
clusters with mass estimated from the dynamical analysis of the Sloan
spectroscopic data.  The filled
points are systems with mass estimated from the $M-T_X$ relation. The
dot-dashed line is the best fit line obtained for the O sample.  The
dashed line is the best fit line for the X sample, while the solid
line is the result obtained from the E sample.}
\label{ML}
\end{figure}

\begin{figure}
\begin{center}
\begin{minipage}{0.5\textwidth}
\resizebox{\hsize}{!}{\includegraphics{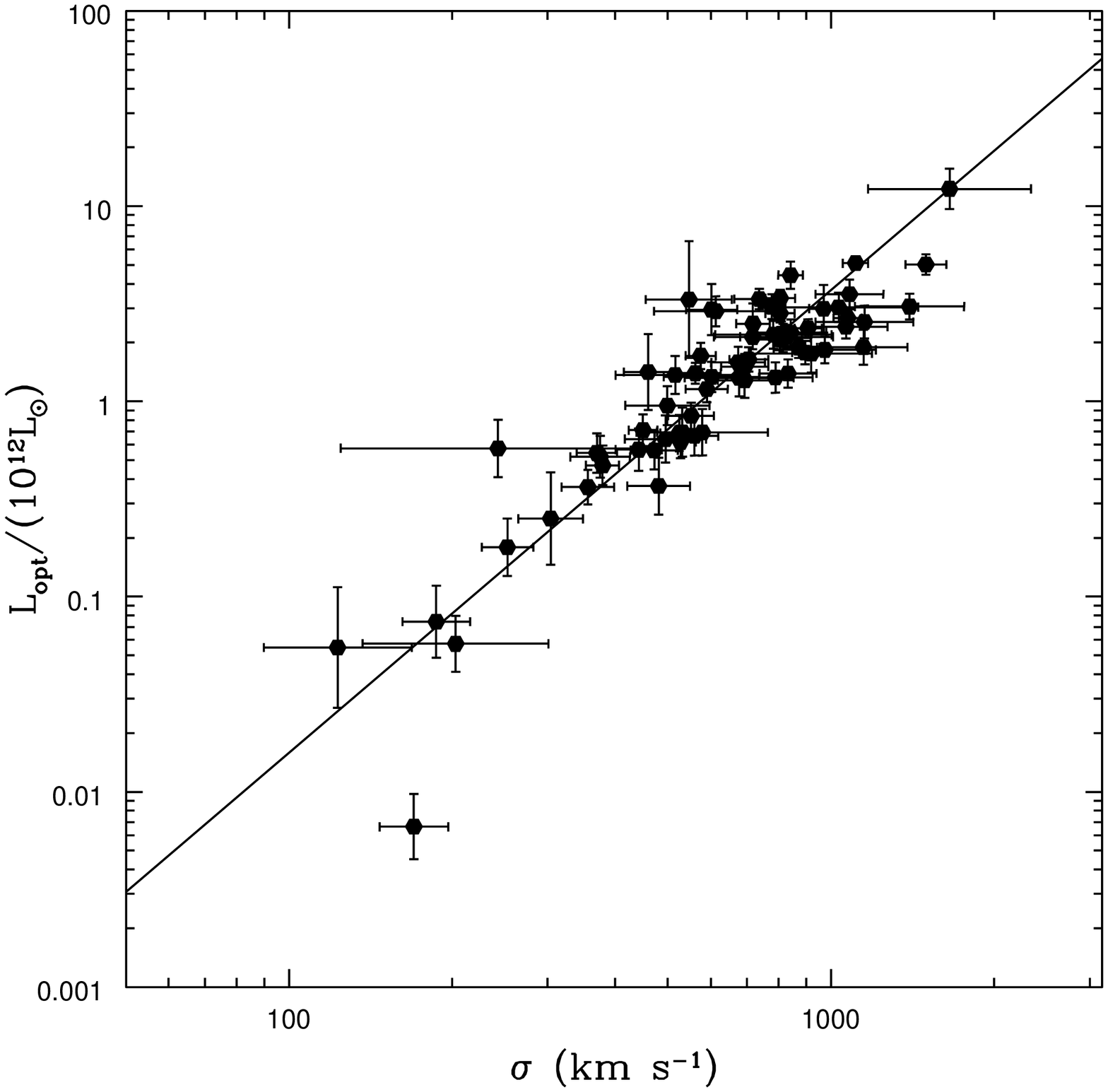}}
\end{minipage}
\end{center}
\caption{  
$L_{op}-\sigma_V$ relation.  The optical luminosities are calculated
within $r_{500}$.  The best-fit line is also shown.}
\label{LO_s}
\end{figure}

\begin{figure}
\begin{center}
\begin{minipage}{0.5\textwidth}
\resizebox{\hsize}{!}{\includegraphics{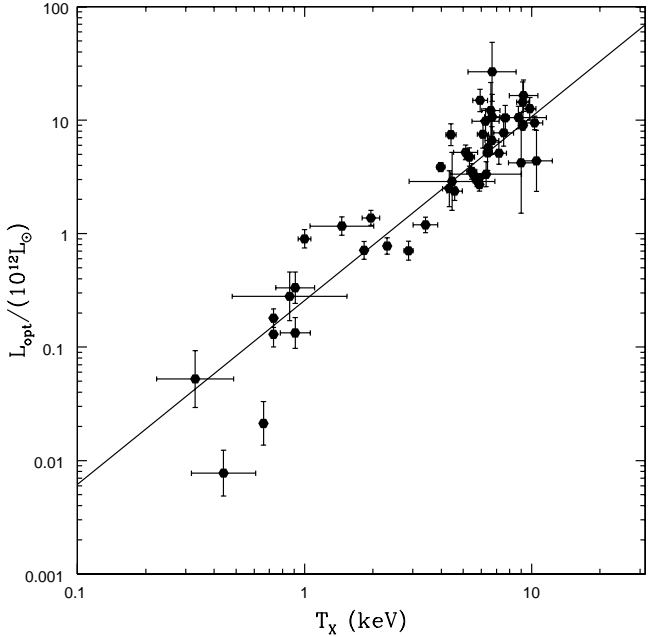}}
\end{minipage}
\end{center}
\caption{
$L_{op}-T_X$ relation.  The optical luminosities are
calculated within $r_{500}$.  
The best-fit line is also shown.}
\label{LO_T}
\end{figure}

The most important and interesting conclusion of this analysis is that
both the optical and the X-ray luminosity show extremely good
relations with the cluster mass within $r_{500}$ and $r_{200}$. The
optical luminosity correlates with the cluster mass better than the
X-ray luminosity and can be used as mass estimator with an average
accuracy of 40\% in the mass determination.  The X-ray luminosity can
predict the cluster mass with an accuracy of 55\% on average. The
presence of a large number of cool core clusters in our sample does
not seem to be the cause of the larger scatter in the $L_X-M$ relation
in comparison to the $L_{op}-M$ relation. 

Our result is in excellent agreement with that obtained by
Lin et al. (2003), who used the $K$-band luminosity as a mass
predictor, and found an average accuracy of 45\%.

On the basis of these results the scatter of the $L_{op}-L_X$ relation
has a natural explanation.  As shown in Table \ref{table1} the values
of all the estimated scatters are very close to the values calculated
for the $L_X-M$ relation.  Therefore, we conclude that the scatter in
the $L_{op}-L_X$ relation (Fig. \ref{LX_LO}) is mostly derived from
the scatter in the $L_X-M$ relation.

\subsection{The Mass-to-Light ratio}
To conclude the analysis of the relation between the optical
luminosity and the cluster mass, we consider the mass-to-light ratio,
M/L, as a function of the cluster mass. 

Previous analyses have shown that, in general, M/L increases with
the cluster mass. Assuming a relation of the type $M/L \propto
M^{\alpha}$, and adopting the usual scaling relations between mass and
X-ray temperature or velocity dispersion, when needed, most authors
have found $\alpha \simeq 0.25 \pm 0.1$, in both optical and
near-infrared bands, and over a very large mass range, from loose
groups to rich clusters of galaxies (Adami et al. 1998; Bahcall \&
Comerford 2002; Girardi et al. 2002; Lin et al. 2003, 2004; Rines et
al. 2004; Ramella et al. 2004; see however Kochanek et al. 2003 for
a discordant result).

Our result is shown in Fig. \ref{ML}, where we plot M/L (in the $i$
band) calculated within $r_{500}$, versus $M_{500}$.  An increase of
the mass-to-light ratio with the mass is clearly visible.  The
existence of a correlation is confirmed by a Spearman's correlation
analysis (the correlation cofficient is 0.42, corresponding to a
probability of only $7\times 10^{-6}$ that the two quantities are not
correlated).  In order to quantify the relation between mass and
luminosity, we prefer to use the $L_{op}-M$ relation directly. In
fact, since M/L is defined as a function of M and L, it is not correct
to search for the best-fitting relation of M/L versus M or L.  The
$L_{op}-M$ relation implies $M/L \propto M^{0.2\pm 0.08}$.  Therefore
the mass-to-light ratio of galaxy clusters is not constant, but
(slightly) increases with the cluster mass. Our relation (derived in
the $i$ band) is clearly consistent with the relations found in other
bands ($B$-band, Girardi et al. 2002; $V$-band, Bahcall \& Comerford
2002; $R$-band, Adami et al. 1998; $K$-band, Lin et al. 2003, 2004;
Rines et al. 2004; Ramella et al. 2004), and, as a matter of fact, we
checked that similar M/L vs. M dependencies are found in the other
bands of the SDSS.

The fact that the M/L vs. M relation is wavelength independent clearly
rules out the explanation provided by Bahcall \& Comerford (2002),
namely that more massive clusters have a larger M/L because because
their galaxies contain more aged stellar populations, on average, than
galaxies members of less massive clusters. The most likely explanation
for this M/L variation with M has been provided by Lin et al. (2003):
the overall star formation efficiency must be a decreasing function of
the cluster mass.

\begin{table*}
\begin{center}
\begin{tabular}[b]{c|c|ccccc}
\hline
&&$\alpha $ & $\beta$ & $\sigma$ &$\sigma _B$ & $\sigma_A$ \\
\hline
$L_{op}-\sigma_{V}$ ($r_{500}) $&red m. &$ 2.26\pm 0.13$ & $-6.04 \pm 0.38$ & 0.06 & 0.07 & 0.16 \\
\hline
&all m. &$ 2.36\pm 0.13$ & $-6.29 \pm 0.37$ & 0.06 & 0.07 & 0.17 \\
\hline
$L_{op}-\sigma_{V}$ ($r_{200}) $& red m. &$ 2.33\pm 0.16$ & $-6.02 \pm 0.45$ & 0.06 & 0.08 & 0.17 \\
\hline
& all m. &$ 2.33\pm 0.15$ & $-6.05 \pm 0.42$ & 0.07 & 0.09 & 0.17 \\
\hline
$L_X-\sigma_{V}$ & red m. &$ 3.60\pm 0.29$ & $-10.22 \pm 0.80$ & 0.07 & 0.09 & 0.38 \\
\hline
& all m. &$ 3.68\pm 0.25$ & $-10.53 \pm 0.80$ & 0.08 & 0.08 & 0.40 \\
\hline
$L_{op}-T_X$ & $r_{500}$ &$ 1.68\pm 0.08$ & $-0.50 \pm 0.06$ & 0.08 & 0.09 & 0.15 \\
\hline
& $r_{200}$& $ 1.66\pm 0.09$ & $-0.41 \pm 0.07$ & 0.09 & 0.12 & 0.17 \\
\hline
$L_X-T_X$ &  &$ 3.06\pm 0.10$ & $-1.77 \pm 0.07$ & 0.07 & 0.10 & 0.29 \\
\hline
\end{tabular}	
\caption{ 
The table lists the best fit values for several correlations:
$L_{op}-\sigma_V$, $L_X-\sigma_V$, $L_{op}-T_X$ and $L_X-T_X$. The
table shows the results obtained with the dynamical analysis performed
on the red members of the systems ('red m.' in the table) and with the
complete cluster membership ('all m'. in the table).  The table
lists three estimation of the scatter for each relation: $\sigma$ is
the orthogonal scatter of the A-B relation, $\sigma_A$ is the scatter
in the A variable and $\sigma_B$ is the scatter in the B variable.
All the scatters in the table are expressed in dex, while all the
errors are given at the 95\% confidence level.}
\label{table2}
\end{center}							   
\end{table*}

\begin{figure}
\begin{center}
\begin{minipage}{0.5\textwidth}
\resizebox{\hsize}{!}{\includegraphics{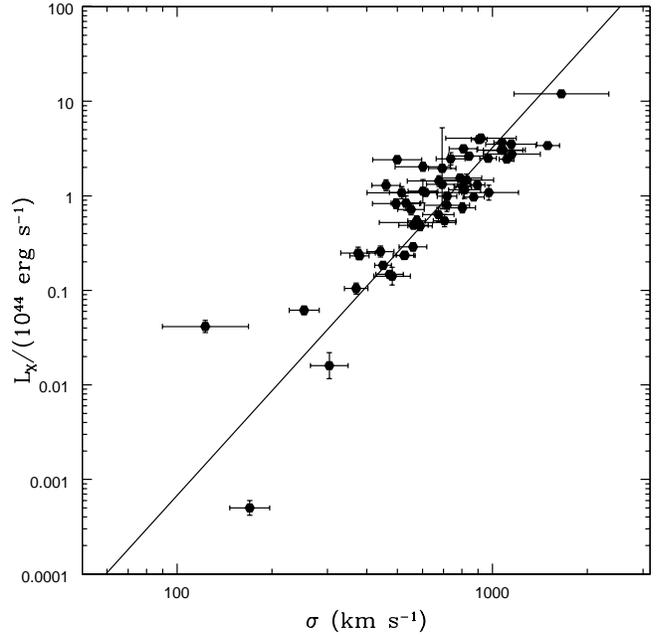}}
\end{minipage}
\end{center}
\caption{  
$L_X-\sigma_V$ relation. 
The best-fit line is also shown.}
\label{LX_s}
\end{figure}

\begin{figure}
\begin{center}
\begin{minipage}{0.5\textwidth}
\resizebox{\hsize}{!}{\includegraphics{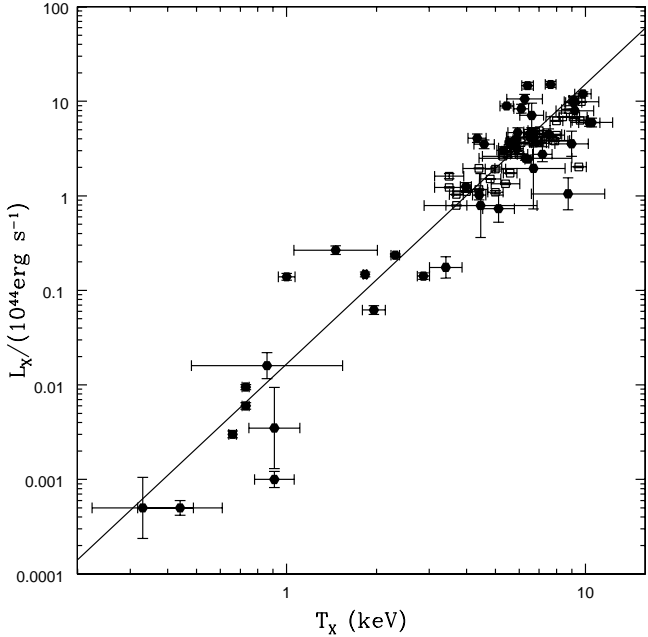}}
\end{minipage}
\end{center}
\caption{ 
$L_X-T_X$ relation. The filled points are the RASS-SDSS clusters with
known ASCA temperature,while the empty squares are the clusters of
M98. The best-fit line is also shown.}
\label{LX_T}
\end{figure}

\begin{figure}
\begin{center}
\begin{minipage}{0.5\textwidth}
\resizebox{\hsize}{!}{\includegraphics{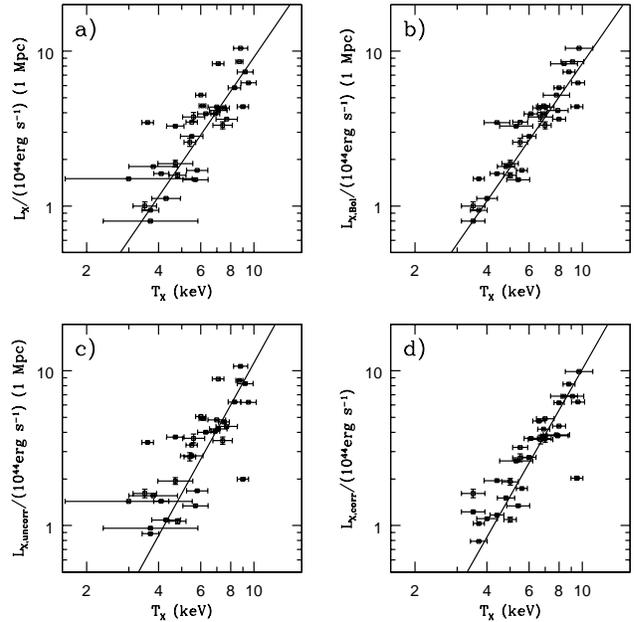}}
\end{minipage}
\end{center}
\caption{
The same as in fig. \ref{Mark} but for the $L_X-T_X$ relation.}
\label{Mark_t}
\end{figure}

\begin{table*}
\begin{center}
\begin{tabular}[b]{c|ccccc}
\hline
\renewcommand{\arraystretch}{0.2}\renewcommand{\tabcolsep}{0.05cm}
& $\alpha $ & $\beta$ & $\sigma$ &$\sigma_A$ & $\sigma_B$ \\
\hline
$L_X(0.1-2.4 keV)-T_X$ (1.4 Mpc,uncorr) & $ 2.59\pm 0.38$ & $-1.48 \pm 0.31$ & 0.06 & 0.07 & 0.18 \\
\hline
$L_X(0.1-2.4 keV)-T_X$ (1.4 Mpc,corr) & $ 2.26\pm 0.19$ & $-1.30 \pm 0.15$ & 0.04 & 0.05 & 0.11 \\
\hline	
$L_{X,corr}(0.1-2.4 keV)-T_X$ (tot,uncorr) & $ 3.30\pm 0.67$ & $-2.06 \pm 0.54$ & 0.06 & 0.09 & 0.24 \\
\hline	
$L_{X,uncorr}(0.1-2.4 keV)-T_X$ (tot,corr) & $ 2.80\pm0.38 $ & $-1.75\pm0.30$ & 0.05 & 0.07 & 0.15 \\
\hline
\end{tabular}	
\caption{ 
The table lists the best fit values for the $L_X-T_X$ relations.  The
first two lines list the relations obtained with the M98 data:
$L_X-T_X$ with $L_X$ calculated in 0.1-2.4 keV energy band, within 1
Mpc from the cluster center and the temperature uncorrected for cool
core effect, $L_X-T_X$ with the X-ray luminosity and temperature
corrected for cool core effect.  The last two lines list the
correlations obtained with the total X-ray luminosity taken from R02
and the temperature of M98 corrected and uncorrected for cool core
effect respectively. All the scatters in the table are expressed in
dex and have the same meaning as in the previous table.}
\label{table4}
\end{center}							   
\end{table*}

\subsection{Correlations of the optical and the X-ray luminosities with the cluster temperature and velocity dispersion.}

The X-ray temperature, $T_X$, and the cluster velocity dispersion,
$\sigma_V$, have both been used as key measures of cluster properties
and in particular of the cluster mass. Given the excellent correlation
of the optical luminosity with the mass, $L_{op}$ should also
correlate with, and have predictive power for, these two quantities.
Table \ref{table2} summarizes the results obtained by correlating the
optical luminosity of the $i$ Sloan band within $r_{500}$ and
$r_{200}$ with $T_X$ and $\sigma_V$.  As shown by Figs. \ref{LO_s} and
\ref{LO_T}, the optical luminosity correlates very well with both
$T_X$ and $\sigma_V$ with an orthogonal scatter of 22\% and 15\%
respectevely. Moreover, $L_{op}$ can predict $T_X$ with 23-28\% accuracy
and $\sigma_V$ with a 17-23\% accuracy. Table \ref{table2} contains the
best fit results also for the $L_X-T_X$ and $L_X-\sigma_V$ relations.
The best fit value of the relations are perfectly in agreement with
the results of Ortiz-Gil et al. (2004), who used a subsample of the
REFLEX sample. The X-ray luminosity defined in the REFLEX catalog are
calculated with the same method used for the RASS-SDSS cluster
catalog.  The orthogonal scatter of the $L_X-\sigma_V$ relation (17\%) is also in
good agreement with Ortiz-Gil et al.  (2004), if we consider the
relation obtained in that work using only clusters with accurate
$\sigma_V$ estimation.  As shown by Figs.
\ref{LX_s} and \ref{LX_T}, also the X-ray luminosity shows a tight
correlation with both quantities and the scatter of the best fit line
(25-30\% accuracy in the $T_X$ prediction and 20-23\% accuracy in
the $\sigma_V$ prediction) is very close to the results obtained
with the optical luminosity . $L_{op}$ is slightly a better
predictor than $L_X$ with a 5\% difference in the scatter.  It is
interesting to notice that, while $L_{op}$ is a much better predictor
of the cluster mass in comparison to $L_X$, optical and X-ray
luminosities can predict approximately with the same accuracy (20\%)
the intracluster temperature and the galaxy velocity dispersion
. This different behavior of the scatter in the relations
involving $L_X$ could be due to the dependence of the X-ray luminosity
and temperature on the cluster compactness. The cluster temperature is
proportional to $M/R$, where $M$ is the cluster mass and $R$ is a
characteristic radius of the system. Thus, $T_X$ is related to mass
with a weighting for compactness. As explained in the previous
paragraph, $L_X$ is proportional to the gas density squared. This implies
that, at given mass, a compact cluster is much more X-ray bright than
a less compact one. Therefore, the cluster compactness could enter the
$L_X-M$ relation as a third parameter, explaining the observed large
scatter. However, since both $L_X$ and $T_X$ have a similar dependence
on the compactness, the dispersion in the $L_X-T_X$ relation would not
be affected. This would explain why $L_{op}$ is a better estimator of
the cluster mass in comparison to $L_X$, while optical and X-ray
luminosities have similar scatter in their relation with $T_X$ and
$\sigma_V$. 

As in the case of the luminosity-mass relation, we investigate in more
details the luminosity-temperature relation to understand which of the
two luminosities is the best predictor of the other cluster
parameters.  We use the X-ray luminosity (calculated within 1.4 Mpc)
and the temperature from M98 to check the influence of the cool-core
correction in the $L_X-T_X$ relation (Fig. \ref{Mark_t}). As shown in
Table \ref{table4}, using the cool-core-corrected X-ray luminosity and
temperature lowers the scatter in the $T_X$ variable by 6\%.  We
analyse the relation using also the total X-ray luminosity of R02 with
and without the cool-core correction (panel $c$ and $d$, respectively,
in Fig. \ref{Mark_t}), as we did for the $L_X-M$ relation.  In both
cases, using the total $L_X$ not only affects the slope of the
relation, but also increases the scatter by 6\% in comparison to the
relations obtained using M98's $L_X$'s.

The slope of the $L_X-T_X$ relation obtained applying the
cool-core-correction to the total luminosity and temperature is
perfectly in agreement with the results obtained previously with the
subsample of 'uncorrected' RASS-SDSS clusters, while the scatter is
lower by 5\%. Such a reduction of the scatter makes the X-ray
luminosity a predictor of the X-ray temperature at least as good as 
the optical luminosity.  However, a similar scatter reduction could in
principle be expected for the $L_{op}-T_X$ relation also, since the
cool-core correction affects X-ray temperatures much more than X-ray
luminosities (M98).  Unfortunately only few of the clusters of the M98
sample are in the sky region covered by the SDSS. Thus, we cannot
check directly this possibility.

\section{Summary and conclusions}
We used the RASS-SDSS galaxy cluster sample to compare the quality of
the optical and X-ray luminosity as predictors of other cluster
properties such as the mass within $r_{500}$ and $r_{200}$, the
velocity dispersion and the ICM temperature. The optical luminosity
turns out to be a better predictor of the cluster mass than the X-ray
luminosity. The knowledge of $L_{op}$ allows to estimate the cluster
mass with an average accuracy of 40\%, while $L_X$ can be used to
predict the mass with an average accuracy of 55\%. We investigated the
nature of the scatter of the $L_X-M$ relation using a sample of
clusters with X-ray luminosity corrected for the effect of cool core
at the center of the system. We concluded that this kind of effect can
affect the scatter of the relation by at most 3\% and, thus, it cannot
explain the dispersion in the observed $L_X-M$ relation, which is
probably related with the variation in the compactness of the
galaxy clusters. We conclude that a cluster optical luminosity is a
better estimator of its mass than its X-ray luminosity. The optical
luminosity is clearly a very useful and rather cheap estimator (in
terms of observational resources required) given that it can be
determined from ground-based photometric data only.

We also analysed the relations of the optical and X-ray luminosities
with cluster velocity dispersions and X-ray temperatures. We found
that both luminosities are strongly correlated with these cluster
properties, and can be used to predict them with an average accuracy
of 20\%. Using a sample of clusters with $L_X$ and $T_X$ corrected for
the cool core effect, we find that the scatter in the $L_X-T_X$
relation is decreased by 5\%. Such a decrease is almost exclusively
due to the correction applied to the X-ray temperature, since the
cool-core correction has a negligible effect on the X-ray
luminosity. Therefore, we expect a similar decrease of the scatter of
the $L_{op}-T_X$ relation, when the X-ray temperatures are corrected
for the same cool-core effect. Unfortunately we cannot verify this
expectation on our sample, since we lack the information to apply the
cool-core correction to the clusters with known $L_{op}$.  We conclude
that $L_{op}$ and $L_X$ can be used to predict the ICM
temperature and the cluster velocity dispersion, at a similar
level of accuracy.

The most important conclusion of our analysis is that the optical
luminosity is a key measure of the fundamental properties of a galaxy
cluster, such as its mass, velocity dispersion, and temperature. In
this respect, the optical luminosity performs even better than the
X-ray luminosity, which suggests that the mass distribution of a
cluster is better traced by cluster galaxies rather than by
intracluster gas (see, e.g., the discussion in Biviano \& Girardi
2003). The poorer performance of $L_X$ as a cluster mass predictor,
relative to $L_{op}$, is probably related to the variation in the compactness of the galaxy clusters.

Our conclusion is clearly in agreement with Lin et al.'s (2003)
result, namely that the $K$-band luminosity is a good estimator of the
cluster mass. On the other hand, our conclusion is at odds with the
generally accepted view that a cluster main physical properties are
more easily revealed in the X-ray than in the optical (e.g.  Donahue
et al. 2002). Such a view was established at an epoch when the lack of
optical wide field surveys precluded a reliable determination of the
optical luminosities of a large sample of clusters. With the advent of
the Sloan Digital Sky survey, this problem is now overcome, and
$L_{op}$ can now be used to infer the fundamental physical properties
of the many clusters being discovered within large optical surveys
with improved cluster finding techniques (such as, e.g., the
Red-Sequence Cluster Survey, Barrientos et al.  2003). In this context
$L_{op}$ becomes a very powerful mean to do cosmological studies with
galaxy clusters without the need of optical or X-ray spectroscopy.

Finally, we showed that the relation between mass and luminosity
implies an increasing mass-to-light ratio, M/L, with increasing
cluster mass. The dependence we found is in excellent agreement with
previous results (Adami et al. 1998; Bahcall \& Comerford 2002;
Girardi et al. 2002; Lin et al. 2003, 2004; Rines et al. 2004; Ramella
et al. 2004), and confirms the achromaticity of the effect. Hence, the
effect cannot be explained by the different ages of the galaxies stellar
populations in clusters of different masses (Bahcall \& Comerford
2002), but, rather, seems to indicate that the star formation
efficiency decreases as the cluster mass increases (Lin et al. 2003).

The results obtained in this paper are applicable to nearby clusters
since the RASS-SDSS galaxy cluster catalog comprises only X-ray
clusters at $z \le 0.3$. It would be very interesting to conduct the
same analysis on a sample of high redshift clusters to study the
evolution of the analysed relations and to compare again the quality
of $L_{op}$ and $L_X$ as predictors of the other cluster properties.

The analysis conducted in this paper is based on a sample of clusters
all detected in the X-ray, 90\% of which are taken from X-ray-selected
galaxy cluster catalogs. It seems that optically bright clusters
exist which are faint in the X-ray (Donahue et al. 2002).  Hence, the
selection criteria of the RASS-SDSS galaxy cluster catalog could in
principle affects our results (see, e.g., Gilbank et al. 2004).  To
check how the selection criteria affect the correlations studied in
this paper, we plan to repeat the same analyses on a sample of
optically-selected clusters.
  
\vspace{2cm}

Funding for the creation and distribution of the SDSS Archive has been
provided  by    the  Alfred P.   Sloan   Foundation, the Participating
Institutions, the  National Aeronautics and  Space Administration, the
National  Science  Foundation, the  U.S.    Department of  Energy, the
Japanese Monbukagakusho, and the Max Planck Society. The SDSS Web site
is   http://www.sdss.org/. The  SDSS is  managed  by the Astrophysical
Research Consortium   (ARC) for the   Participating Institutions.  The
Participating Institutions  are The  University of Chicago,  Fermilab,
the Institute for Advanced  Study, the Japan Participation  Group, The
Johns  Hopkins    University,  Los Alamos   National  Laboratory,  the
Max-Planck-Institute for   Astronomy (MPIA), the  Max-Planck-Institute
for   Astrophysics (MPA), New Mexico  State  University, University of
Pittsburgh, Princeton University, the United States Naval Observatory,
and the University of Washington.

\begin{appendix}

\section{Correlation of $L_{op}$ in the Sloan $g, r$ and $z$ bands with the cluster parameters.}

We list in the table all the results obtained using the optical
luminosity calculated in the g, r and z SDSS Sloan bands. The structure
of the three tables in this appendix is similar to the tables in the
text of the paper. For each analysed correlation we report the value
of the best fit parameters plus the error at 95\% confidence level and
three values of the scatter: the orthogonal scatter of the relation
and the scatters in both variables (in the logarithmic space). All the
scatters are expressed in dex. All the results are obtained using the
mass and velocity dispersion derived from the dynamical analysis of
the red members of the clusters.

\clearpage
\begin{table*}
\begin{minipage}{0.5\textwidth}
\begin{sideways}
\begin{tabular}[b]{c|ccccc|ccccc|ccccc}
\hline
\multicolumn{1}{c}{ }&\multicolumn{5}{c|}{g band}& \multicolumn{5}{|c}{r band}& \multicolumn{5}{|c}{z band} \\ \hline
\multicolumn{16}{c}{$L_{op}-M_{500}$} \\ \hline
\renewcommand{\arraystretch}{0.2}\renewcommand{\tabcolsep}{0.05cm}
&$\alpha $ & $\beta$ & $\sigma$ &$\sigma _{M_{500}}$ & $\sigma_{L_{op}}$ & $\alpha $ & $\beta$ &$\sigma$& $\sigma _{M_{500}}$ & $\sigma_{L_{op}}$ & $\alpha $ & $\beta$ &$\sigma$& $\sigma _{M_{500}}$ & $\sigma_{L_{op}}$\\
\hline
O   &  $  0.81\pm0.05$ & $-0.30\pm 0.03$ & 0.12 & 0.18 & 0.16    &  $  0.80\pm0.04$ & $-0.22\pm 0.03$ & 0.11 & 0.16 & 0.16    &  $  0.91\pm0.04$ & $ 0.00\pm 0.03$ & 0.13 & 0.18 & 0.19 \\ 
X   &  $  1.06\pm0.06$ & $-0.26\pm 0.04$ & 0.14 & 0.22 & 0.17    &  $  1.08\pm0.06$ & $-0.18\pm 0.04$ & 0.12 & 0.18 & 0.16    &  $  1.04\pm0.05$ & $ 0.07\pm 0.04$ & 0.10 & 0.16 & 0.14 \\ 
E   &  $  0.90\pm0.04$ & $-0.31\pm 0.03$ & 0.15 & 0.21 & 0.20    &  $  0.92\pm0.04$ & $-0.24\pm 0.03$ & 0.13 & 0.20 & 0.20    &  $  0.79\pm0.04$ & $ 0.01\pm 0.03$ & 0.10 & 0.16 & 0.15 \\ 
\hline
\multicolumn{16}{c}{$L_{op}-M_{200}$} \\ \hline
\renewcommand{\arraystretch}{0.2}\renewcommand{\tabcolsep}{0.05cm}
&$\alpha $ & $\beta$ & $\sigma$ &$\sigma _{M_{200}}$ & $\sigma_{L_{op}}$ & $\alpha $ & $\beta$ &$\sigma$& $\sigma _{M_{200}}$ & $\sigma_{L_{op}}$ & $\alpha $ & $\beta$ &$\sigma$& $\sigma _{M_{200}}$ & $\sigma_{L_{op}}$\\
\hline
O   &  $  0.81\pm0.04$ & $-0.27\pm 0.04$ & 0.13 & 0.18 & 0.15    &  $  0.80\pm0.04$ & $-0.18\pm 0.04$ & 0.12 & 0.17 & 0.14    &  $  0.79\pm0.04$ & $ 0.06\pm 0.04$ & 0.12 & 0.16 & 0.14 \\ 
X   &  $  1.08\pm0.08$ & $-0.25\pm 0.06$ & 0.14 & 0.23 & 0.20    &  $  1.05\pm0.07$ & $-0.17\pm 0.06$ & 0.12 & 0.20 & 0.17    &  $  1.02\pm0.06$ & $ 0.08\pm 0.06$ & 0.10 & 0.18 & 0.16 \\ 
E   &  $  0.94\pm0.05$ & $-0.31\pm 0.04$ & 0.16 & 0.23 & 0.20    &  $  0.94\pm0.04$ & $-0.24\pm 0.04$ & 0.15 & 0.21 & 0.19    &  $  0.93\pm0.04$ & $ 0.00\pm 0.04$ & 0.14 & 0.21 & 0.19 \\ 
\hline
\multicolumn{16}{c}{$L_{op}-L_X$ ($r_{500}$)} \\ \hline
\renewcommand{\arraystretch}{0.2}\renewcommand{\tabcolsep}{0.05cm}
&$\alpha $ & $\beta$ & $\sigma$ &$\sigma _{L_X}$ & $\sigma_{L_{op}}$ & $\alpha $ & $\beta$ &$\sigma$& $\sigma _{L_X}$ & $\sigma_{L_{op}}$ & $\alpha $ & $\beta$ &$\sigma$& $\sigma _{L_X}$ & $\sigma_{L_{op}}$\\
\hline
O   &  $  0.62\pm0.05$ & $ 0.21\pm 0.02$ & 0.15 & 0.31 & 0.18    &  $  0.63\pm0.04$ & $ 0.28\pm 0.02$ & 0.14 & 0.29 & 0.17    &  $  0.63\pm0.04$ & $ 0.52\pm 0.02$ & 0.13 & 0.28 & 0.16 \\ 
X   &  $  0.51\pm0.03$ & $ 0.19\pm 0.03$ & 0.19 & 0.38 & 0.20    &  $  0.53\pm0.03$ & $ 0.27\pm 0.03$ & 0.18 & 0.33 & 0.20    &  $  0.52\pm0.03$ & $ 0.49\pm 0.03$ & 0.15 & 0.32 & 0.18 \\ 
E   &  $  0.55\pm0.03$ & $ 0.21\pm 0.02$ & 0.17 & 0.35 & 0.19    &  $  0.56\pm0.03$ & $ 0.28\pm 0.02$ & 0.15 & 0.32 & 0.18    &  $  0.56\pm0.03$ & $ 0.51\pm 0.02$ & 0.14 & 0.31 & 0.17 \\ 
\hline
\multicolumn{16}{c}{$L_{op}-L_X$ ($r_{200}$)} \\ \hline
\renewcommand{\arraystretch}{0.2}\renewcommand{\tabcolsep}{0.05cm}
&$\alpha $ & $\beta$ & $\sigma$ &$\sigma _{L_X}$ & $\sigma_{L_{op}}$ & $\alpha $ & $\beta$ &$\sigma$& $\sigma _{L_X}$ & $\sigma_{L_{op}}$ & $\alpha $ & $\beta$ &$\sigma$& $\sigma _{L_X}$ & $\sigma_{L_{op}}$\\
\hline
O   &  $  0.63\pm0.05$ & $ 0.37\pm 0.02$ & 0.17 & 0.28 & 0.18    &  $  0.64\pm0.04$ & $ 0.45\pm 0.02$ & 0.15 & 0.26 & 0.17    &  $  0.64\pm0.04$ & $ 0.68\pm 0.02$ & 0.15 & 0.25 & 0.16 \\ 
X   &  $  0.54\pm0.04$ & $ 0.37\pm 0.03$ & 0.22 & 0.39 & 0.23    &  $  0.54\pm0.04$ & $ 0.44\pm 0.03$ & 0.20 & 0.36 & 0.22    &  $  0.54\pm0.04$ & $ 0.67\pm 0.03$ & 0.19 & 0.35 & 0.21 \\ 
E   &  $  0.57\pm0.03$ & $ 0.37\pm 0.02$ & 0.19 & 0.34 & 0.19    &  $  0.58\pm0.03$ & $ 0.45\pm 0.02$ & 0.17 & 0.31 & 0.19    &  $  0.58\pm0.03$ & $ 0.68\pm 0.02$ & 0.16 & 0.30 & 0.18 \\ 
\hline
\multicolumn{16}{c}{$L_{op}-r_{500}$} \\ \hline
\renewcommand{\arraystretch}{0.2}\renewcommand{\tabcolsep}{0.05cm}
&$\alpha $ & $\beta$ & $\sigma$ &$\sigma _{r_{500}}$ & $\sigma_{L_{op}}$ & $\alpha $ & $\beta$ &$\sigma$& $\sigma _{r_{500}}$ & $\sigma_{L_{op}}$ & $\alpha $ & $\beta$ &$\sigma$& $\sigma _{r_{500}}$ & $\sigma_{L_{op}}$\\
\hline
O&  $2.24\pm0.15$ & $0.11\pm 0.02$ &  0.06&  0.06& 0.16 &  $2.24\pm0.14$ & $0.19\pm0.02$ & 0.05& 0.06& 0.15&  $2.27\pm0.14$ & $0.42\pm 0.02$& 0.05& 0.05 & 0.15 \\
X&  $2.95\pm0.16$ & $0.23\pm 0.03$ &  0.06&  0.07& 0.17 &  $2.97\pm0.16$ & $0.31\pm0.02$ & 0.05& 0.06& 0.15&  $2.93\pm0.15$ & $0.54\pm 0.02$& 0.04& 0.05 & 0.14 \\
E&  $2.44\pm0.12$ & $0.14\pm 0.02$ &  0.07&  0.09& 0.18 &  $2.50\pm0.12$ & $0.23\pm0.02$ & 0.06& 0.08& 0.18&  $2.50\pm0.12$ & $0.47\pm 0.02$& 0.06& 0.07 & 0.17 \\ 
\hline
\multicolumn{16}{c}{$L_{op}-r_{200}$} \\ \hline
\renewcommand{\arraystretch}{0.2}\renewcommand{\tabcolsep}{0.05cm}
&$\alpha $ & $\beta$ & $\sigma$ &$\sigma _{r_{200}}$ & $\sigma_{L_{op}}$ & $\alpha $ & $\beta$ &$\sigma$& $\sigma _{r_{200}}$ & $\sigma_{L_{op}}$ & $\alpha $ & $\beta$ &$\sigma$& $\sigma _{r_{200}}$ & $\sigma_{L_{op}}$\\
\hline
O&  $2.27\pm0.15$ & $-0.14\pm0.04$ &  0.07 & 0.07& 0.15 &  $2.24\pm0.13$ & $-0.06\pm0.04$ &  0.06 & 0.07 &  0.14 & $2.26\pm  0.13$ &  $0.17\pm 0.04$ &  0.06&  0.06 &  0.14\\
X&  $2.27\pm0.15$ & $-0.14\pm0.04$ &  0.07 & 0.07& 0.15 &  $2.85\pm0.18$ & $-0.05\pm0.05$ &  0.06 & 0.09 &  0.17 & $2.81\pm  0.18$ &  $0.19\pm 0.05$ &  0.06&  0.08 &  0.16\\
E&  $2.44\pm0.13$ & $-0.14\pm0.04$ &  0.08 & 0.11& 0.18 &  $2.47\pm0.12$ & $-0.06\pm0.04$ &  0.07 & 0.09 &  0.17 & $2.45\pm  0.12$ &  $0.18\pm 0.04$ &  0.07&  0.09 &  0.17\\
\hline
\multicolumn{16}{c}{$L_{op}-{\sigma_V}$} \\ \hline
\renewcommand{\arraystretch}{0.2}\renewcommand{\tabcolsep}{0.05cm}
&$\alpha $ & $\beta$ & $\sigma$ &$\sigma _{\sigma_V}$ & $\sigma_{L_{op}}$ & $\alpha $ & $\beta$ &$\sigma$& $\sigma _{\sigma_V}$ & $\sigma_{L_{op}}$ & $\alpha $ & $\beta$ &$\sigma$& $\sigma _{\sigma_V}$ & $\sigma_{L_{op}}$\\
\hline
$r_{500}$ &$ 2.37\pm0.15$ &$-6.54\pm0.41$ & 0.06  &0.07  &0.18 &  $2.35\pm 0.14$ &$-6.39\pm0.39$  & 0.06&  0.07 & 0.17  & $2.35\pm0.13$ & $-6.17\pm0.38$ &0.06 & 0.06 & 0.17 \\      
$r_{200}$  &$ 2.38\pm0.16$ &$-6.39\pm0.46$ & 0.07  &0.09  &0.18 &  $2.29\pm 0.15$ &$-6.06\pm0.41$  & 0.07&  0.08 & 0.16  & $2.29\pm0.15$ & $-5.84\pm0.42$ &0.07 & 0.08 & 0.17 \\
\hline
\multicolumn{16}{c}{$L_{op}-T_X$} \\ \hline
\renewcommand{\arraystretch}{0.2}\renewcommand{\tabcolsep}{0.05cm}
&$\alpha $ & $\beta$ & $\sigma$ &$\sigma _{T_X}$ & $\sigma_{L_{op}}$ & $\alpha $ & $\beta$ &$\sigma$& $\sigma _{T_X}$ & $\sigma_{L_{op}}$ &$\alpha $ & $\beta$ & $\sigma$ &$\sigma _{T_X}$ & $\sigma_{L_{op}}$ \\
\hline
$r_{500}$  &$ 1.66\pm0.08$ &$-0.82\pm0.06$ & 0.10  &0.09  &0.17 &  $1.69\pm 0.08$ &$-0.77\pm0.06$  & 0.09&  0.08 & 0.16  & $1.65\pm0.08$ & $-0.50\pm0.06$ &0.08 & 0.08 & 0.14 \\
$r_{200}$  &$ 1.62\pm0.10$ &$-0.59\pm0.08$ & 0.12  &0.13  &0.20 &  $1.64\pm 0.09$ &$-0.53\pm0.07$  & 0.10&  0.12 & 0.17  & $1.62\pm0.10$ & $-0.28\pm0.07$ &0.09 & 0.12 & 0.16 \\
\hline
\end{tabular}
\end{sideways}	
\begin{sideways}
\label{res}
\end{sideways}	
\end{minipage}
\end{table*}

\end{appendix}

\end{document}